\documentclass[a4paper,10pt,twoside]{cpc-hepnp}
\usepackage{multicol}
\usepackage{graphicx}
\usepackage{booktabs}
\usepackage{amssymb,bm,mathrsfs,bbm,amscd}
\usepackage[tbtags]{amsmath}
\usepackage{lastpage}

\begin{document}


\fancyhead[c]{\small Submitted to ¡®Chinese Physics C¡¯}
\fancyfoot[C]{\small 010201-\thepage}

\footnotetext[0]{Received 31 June 2015}

\title{Track segment finding with CGEM-IT and matching to tracks in ODC$^*$ }

\footnotetext[0]{$^*$Supported by National Key Basic Research Program of China (2015CB856706), National Natural Science Foundation of China (11575222, 11205184, 11205182, 11121092, 11475185) and Joint Funds of National Natural Science Foundation of China (U1232201)}

\author{%
      Xin-Hua Sun(Ëïлª)$^{1;1)}$\email{sunxh@ihep.ac.cn}%
\quad Liang-Liang Wang(ÍõÁÁÁÁ)$^{1;2)}$\email{llwang@ihep.ac.cn}
\quad Ling-Hui Wu(ÎéÁé»Û)$^{1}$\\
\quad Xu-Dong Ju(¾ÏÐñ¶«)$^{1}$
\quad Qing-Lei Xiu(ÐÞÇàÀÚ)$^{1}$
\quad Liao-Yuan Dong(¶­ÁÇÔ­)$^{1}$\\
\quad Ming-Yi Dong(¶­Ã÷Òå)$^{1}$
\quad Wei-Dong Li(ÀîÎÀ¶«)$^{1}$
\quad Wei-Guo Li(ÀîÎÀ¹ú)$^{1}$\\
\quad Huai-Min Liu(Áõ»³Ãñ)$^{1}$
\quad Qun Ou-Yang(Å·ÑôȺ)$^{1}$
\quad Ye Yuan(Ô¬Ò°)$^{1}$
\quad Yao Zhang(ÕÅÑþ)$^{1}$
}
\maketitle

\address{%
$^1$ Institution of High Energy Physics, Chinese Academy of Sciences, Beijing 100049, China\\
}

\begin{abstract}
	The relative differences in coordinates of Cylindrical-Gas-Electron-Multiplier-Detector-based Inner Tracker (CGEM-IT) clusters are studied to search for track segments in CGEM-IT. With the full simulation of single muon track samples, clear patterns are found and parameterized for the correct cluster combinations. The cluster combinations satisfying the patterns are selected as track segment candidates in CGEM-IT with an efficiency higher than 99\%. The parameters of the track segments are obtained by a helix fitting. Some $\chi^2$ quantities, evaluating the differences in track parameters between the track segments in CGEM-IT and the tracks found in Outer-Drift-Chamber (ODC), are calculated and used to match them. Proper $\chi^2$ requirements are determined as a function of transverse momentum and the matching efficiency is found reasonable.

\end{abstract}



\footnotetext[0]{\hspace*{-3mm}\raisebox{0.3ex}{
$\scriptstyle\copyright$}2013 Institute of High Energy Physics of the Chinese Academy of Sciences}%

\begin{multicols}{2}

\section{Introduction}

The Inner-Drift-Chamber (IDC) of the Beijing Spectrometer III (BESIII)~\cite{BESIII} showed an aging effect gradually. The hit efficiency and spatial resolution will deteriorate and will affect the reconstruction of charged tracks in the future. A Cylindrical-Gas-Electron-Multiplier-Detector-based~\cite{GEM_intro} Inner Tracker (CGEM-IT) was proposed as a candidate for the upgrade of the IDC~\cite{CGEM_CDR}.

The proposed CGEM-IT consists of three layers of CGEM detector. Each CGEM comprises one cathode, three GEM foils and one anode. There are two kinds of strips on the anode readout plane~\cite{anode_readout} which point to two different directions and are used to measure the induced charge from charged particles. The GEANT4-based~\cite{geant4} full simulation package was implemented for CGEM-IT which includes the detailed detector description and a simplified digitization~\cite{CGEM_ClusterKal}. The cluster reconstruction algorithm was developed, in which the continuous fired strips are reconstructed as clusters and the positions are calculated with the charge centroid method~\cite{CGEM_ClusterKal}. The track fitting algorithm for Drift Chamber based on the Kalman Filter method~\cite{MDC_kal} was also extended to incorporate both CGEM-IT clusters and Outer-Drift-Chamber (ODC) hits, so that the spatial and momentum resolutions can be studied for the tracks reconstructed within the tracking system of the CGEM-IT and the ODC~\cite{CGEM_ClusterKal}. This paper describes the track segment finding with CGEM-IT and the track matching between CGEM-IT and ODC.

\section{Track segment finding with CGEM-IT}

As the CGEM-IT includes three layers, usually three clusters can be reconstructed when a charged particle goes through the CGEM-IT in a magnetic field. The two coordinates of the reconstructed clusters are the azimuthal angle $\phi$ and the position $Z$ along the beam direction, where the center of the detector is taken as the origin point. Figure~\ref{fig_phi} shows an example of a charged track and the induced three clusters on the transverse plane. The selection of the correct three-cluster-combinations is called track segment finding with CGEM-IT. The relation of the three clusters from one track can be studied with the relative differences in $\phi$ and $Z$ between the first or the third layer and the second layer. Then each combination of three clusters can be shown as one point on a two-dimensional plane of $\delta\phi$, where the two dimensions are $\delta\phi_{21}=\phi_1-\phi_2$ and  $\delta\phi_{23}=\phi_3-\phi_2$ (the subscript indexes 1, 2, 3 indicate the layer number). Similarly the combinations of three clusters can also be shown on a two-dimensional plane in $\delta Z$ ($\delta Z_{21}=Z_1-Z_2$, $\delta Z_{23}=Z_3-Z_2$).
\begin{center}
\scalebox{0.4}{\includegraphics*[0.0in,3.0in][12.0in,11.0in]{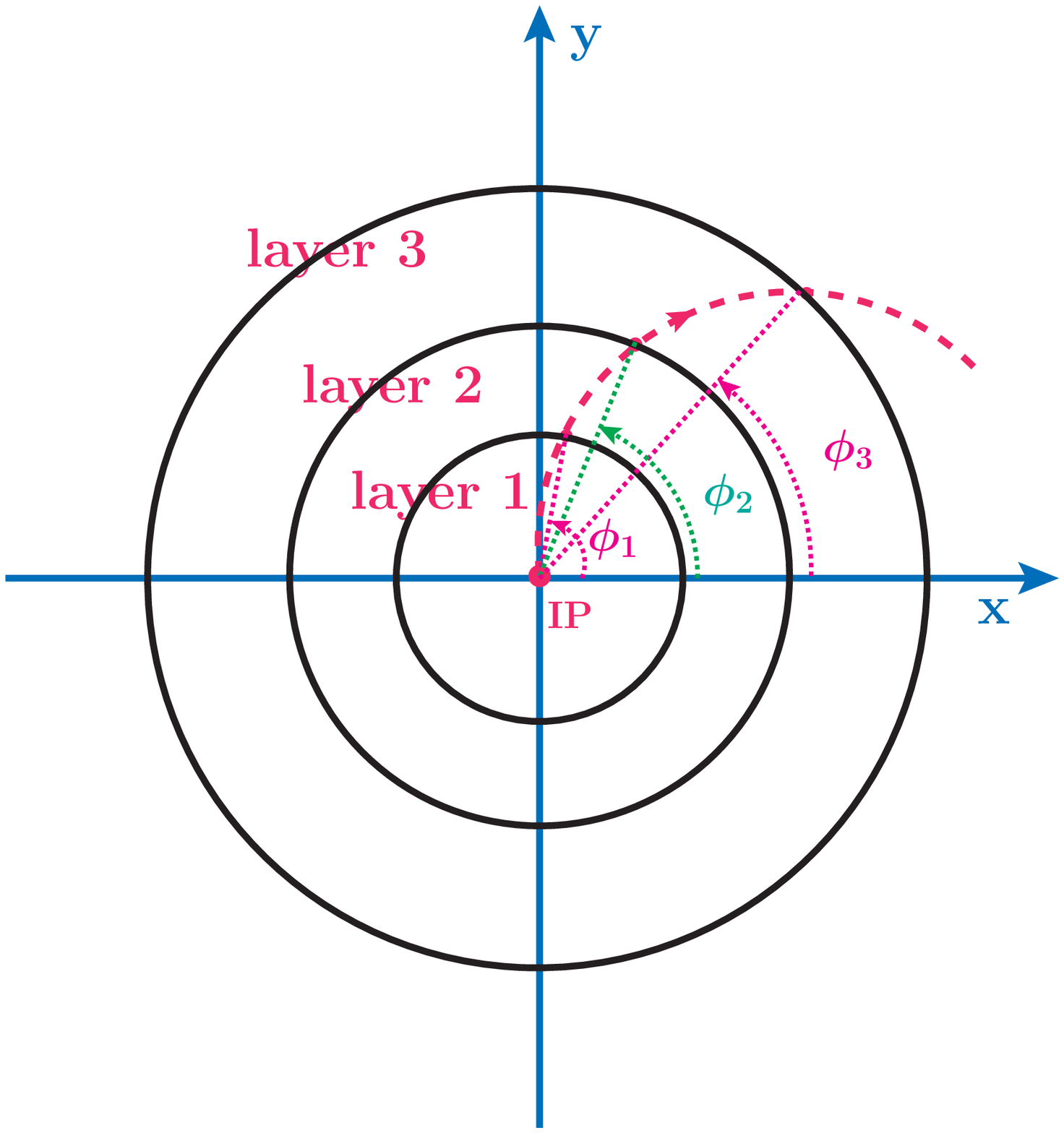}}
\figcaption{\label{fig_phi} (color online) The projection on the transverse plane (perpendicular to beam direction) of the three layers (black circles) of CGEM, and a charged track (red dash line). The three intersections between the CGEM layers and the track are called clusters, the azimuth angles of which are $\phi_1$, $\phi_2$ and $\phi_3$ respectively.}
\end{center}

\subsection{Pattern in $\delta$$\phi$}

Based on the full simulation, the correct three-cluster-combinations on the $\delta\phi$ plane for $\mu^{\pm}$ with a uniform distributed transverse momentum $p_t$ between $0.05$ and $1.0~$GeV$/c$ \footnote{The behavior of higher momentum charged tracks are similar, so the momentum of the charged tracks under study are limited up to 1GeV/c.}, and with a polar angle $\theta$ satisfying $|\cos\theta|<0.93$, are shown in Fig.~\ref{fig2}. A clear anti-diagonal pattern can be observed, the general behavior of which is fitted to a linear function through the origin point. As shown in Fig.~\ref{fig_d_dL}, the scattering points on the $\delta\phi$ plane can be presented by a signed perpendicular distance $d_{\delta\phi}$ to the fitted line (negative for points bellow the line, positive for points above the line) and the projected position $\delta L_{\delta\phi}$ on the line. The $d_{\delta\phi}$ distributions at different $\delta L_{\delta\phi}$ are studied and fitted to a Gaussian function to extract the mean value $\mu_{\delta\phi}$ and the resolution $\sigma_{\delta\phi}$. One example of the $d_{\delta\phi}$ distribution for $0.054<\delta L_{\delta\phi}<0.108$ rad is shown in Fig.~\ref{fig3}.
The obtained mean and resolution values at different $\delta L_{\delta\phi}$ are shown in Fig.~\ref{fig4} which can be fitted to a third order polynomial function without even terms and a fourth order polynomial function without odd terms respectively.
According to the parameterized pattern, the central line of the pattern and a selection window corresponding to a range of $\pm 3\sigma_{\delta\phi}$ for the track segment candidates on the two dimensional $\delta\phi$ plane are determined and shown in Fig.~\ref{fig2}.
\begin{center}
\includegraphics[width=0.4\textwidth]{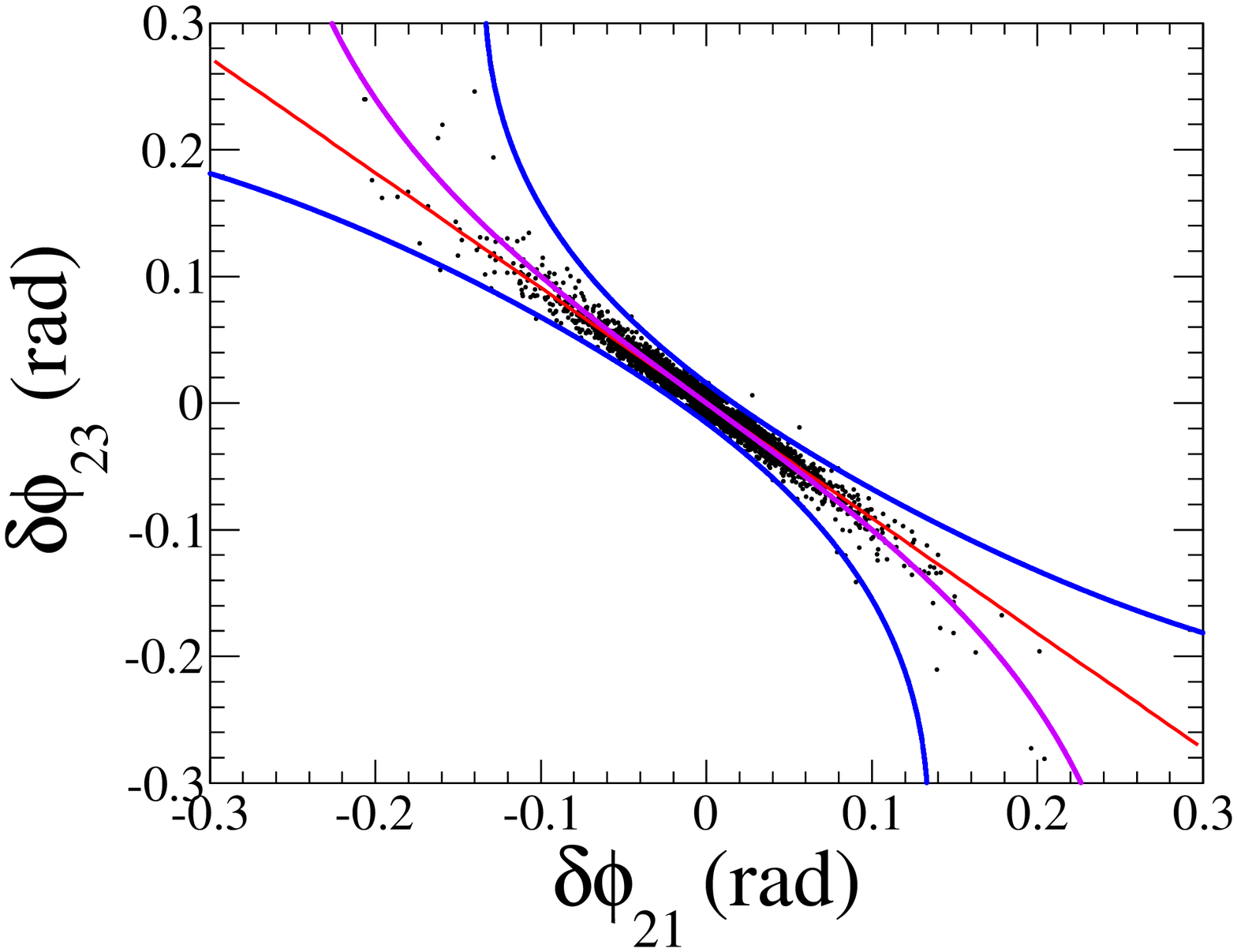}
\figcaption{\label{fig2} (color online) The distribution of correct three-cluster-combinations (black points) on the $\delta\phi$ plane for $\mu^{\pm}$ with random $p_t$ ranging from $0.05$ to $1.0~$GeV$/c$. The red line is the fitted result of the distribution to a linear function through the origin point. The purple curve is the central line of the found pattern which is obtained by fitting the residual distributions along the linear line. The area between the two blue curves is the track segment selection window on the $\delta\phi$ plane which corresponds to a range of $\pm 3\sigma_{\delta\phi}$.}
\end{center}

\begin{center}
\scalebox{0.4}{\includegraphics*[0.0in,5.0in][12.0in,11.0in]{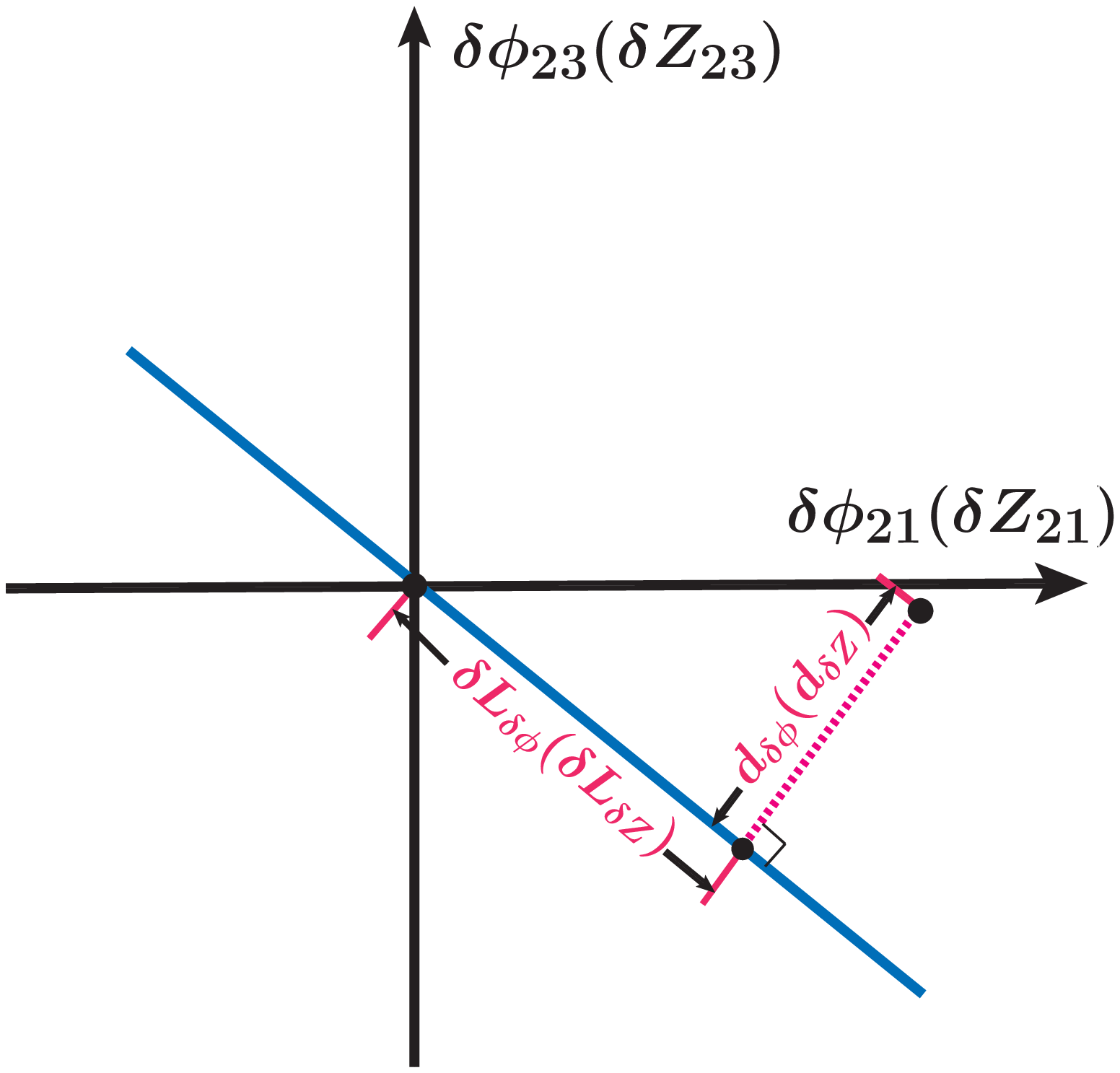}}
\figcaption{\label{fig_d_dL} The definitions of the signed perpendicular distance $d_{\delta\phi}(d_{\delta Z})$ and the projected position $\delta L_{\delta\phi}(\delta L_{\delta Z})$. The black dot is a certain three-cluster-combination and the line is the fitted result described in the text.}
\end{center}
\begin{center}
\includegraphics[width=0.4\textwidth]{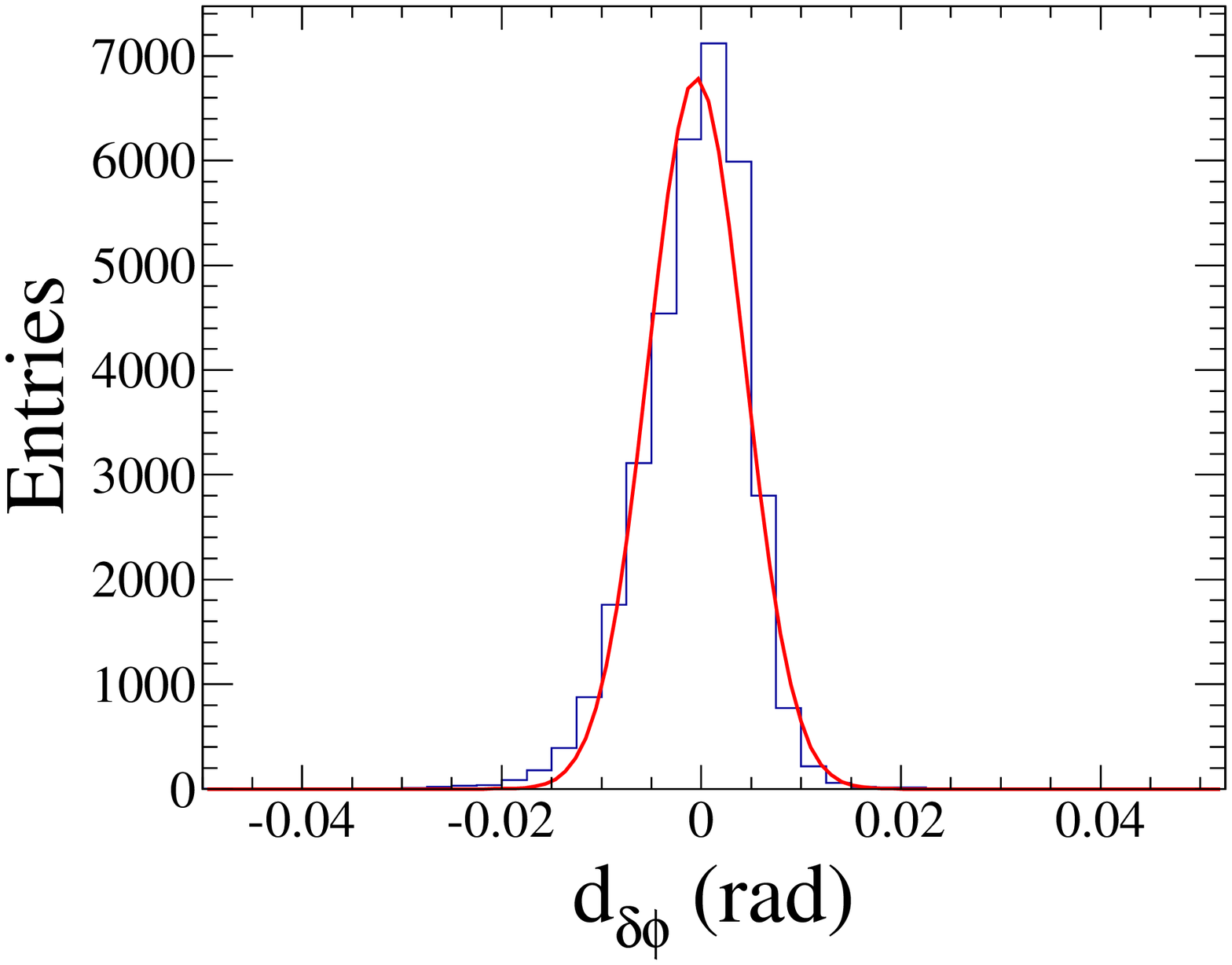}
\figcaption{\label{fig3} The $d_{\delta\phi}$ distribution for $0.054<\delta L_{\delta\phi}<0.108$ rad. The curve is the fitted result to a Gaussian function.}
\end{center}

\begin{center}
\includegraphics[width=0.4\textwidth]{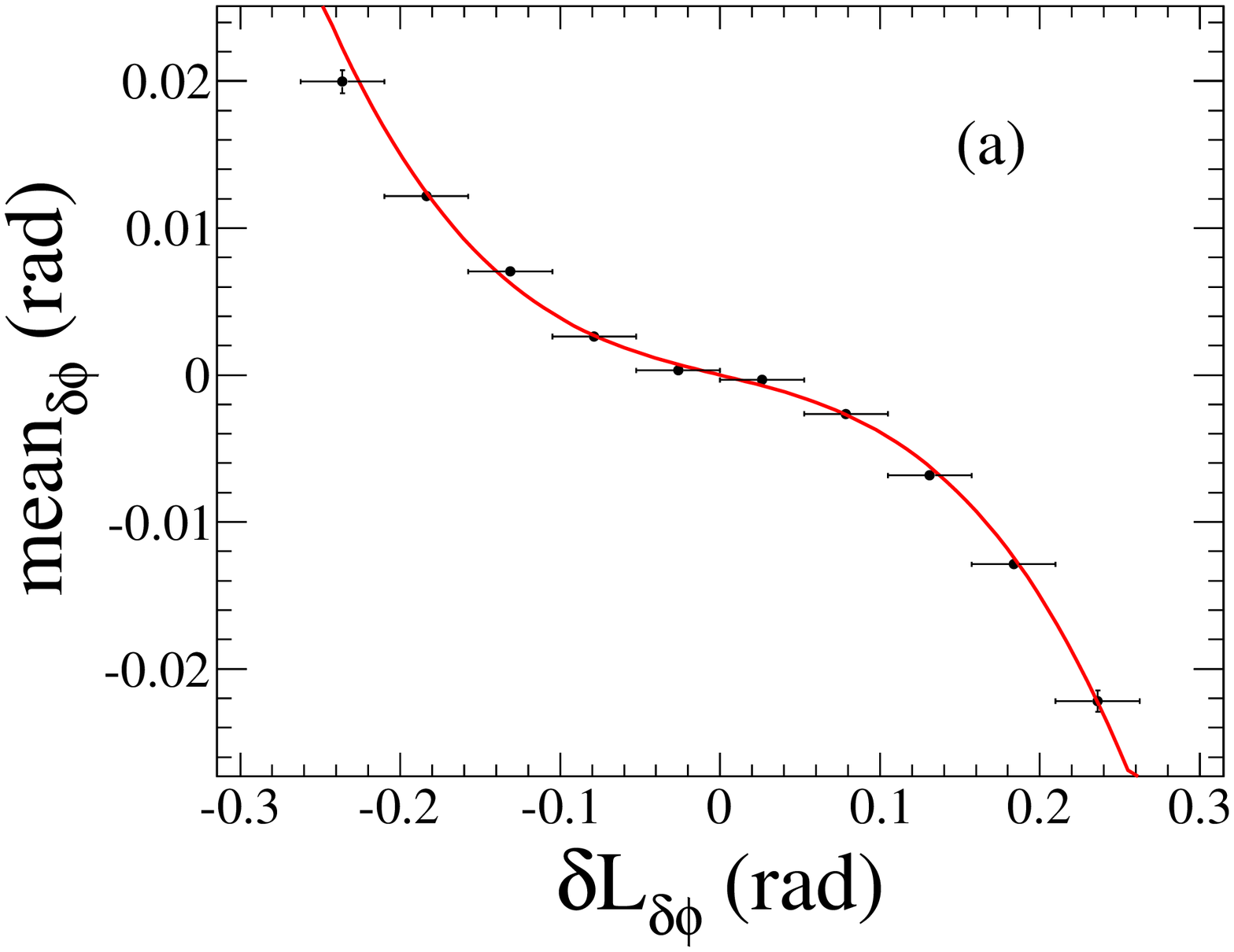}
\includegraphics[width=0.4\textwidth]{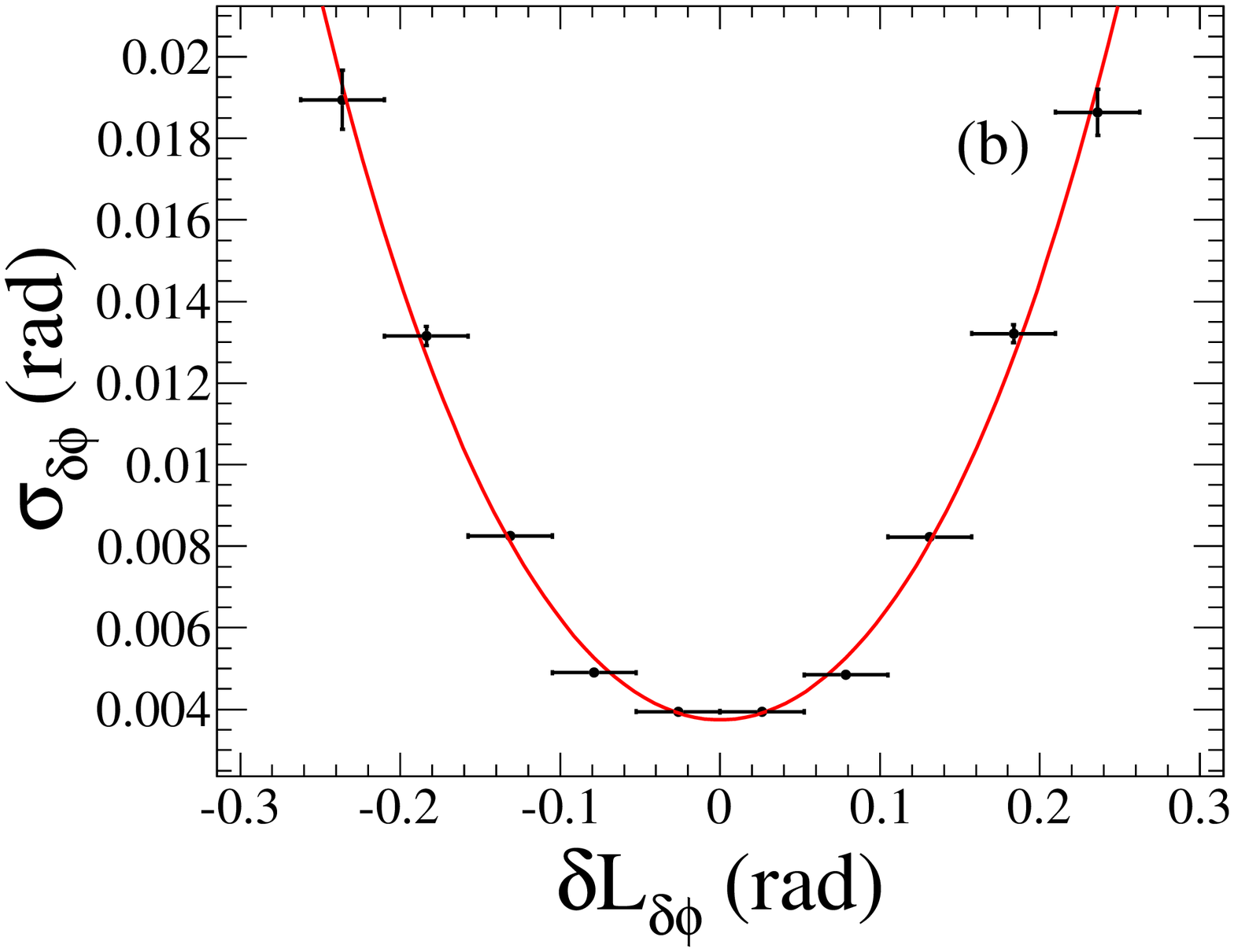}
\figcaption{\label{fig4} The mean (a) of $d_{\delta\phi}$ as a function of $\delta L_{\delta\phi}$ on the $\delta\phi$ plane is fitted to a third order polynomial without even terms (red line). The resolution $\sigma_{\delta\phi}$ (b) of $d_{\delta\phi}$ as a function of $\delta L_{\delta\phi}$ is fitted to a fourth order polynomial without odd terms (red line).}
\end{center}

\subsection{Pattern in $\delta Z$}

\begin{center}
\includegraphics[width=0.4\textwidth]{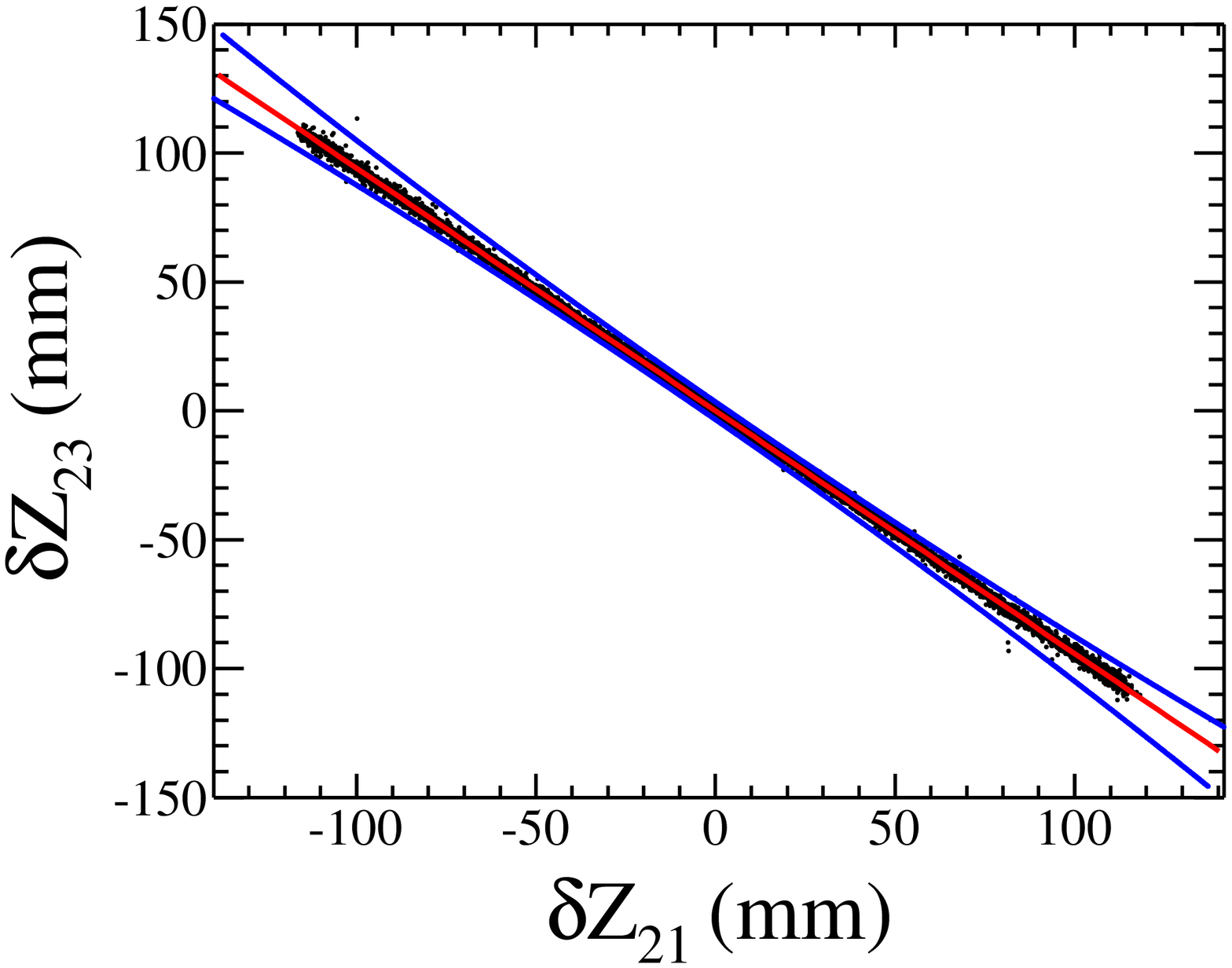}
\figcaption{\label{fig5} (color online)
	The distribution of correct three-cluster-combinations (black points) on the $\delta Z$ plane for $\mu^{\pm}$ with  $p_{t}$ between 0.15 and 0.25 GeV$/c$. The red line is fitted result of the distribution to a linear function through the origin point. The area between two blue lines are the merged three-cluster-combination selection window on the $\delta Z$ plane as described in the text.}
\end{center}
\begin{center}
\includegraphics[width=0.4\textwidth]{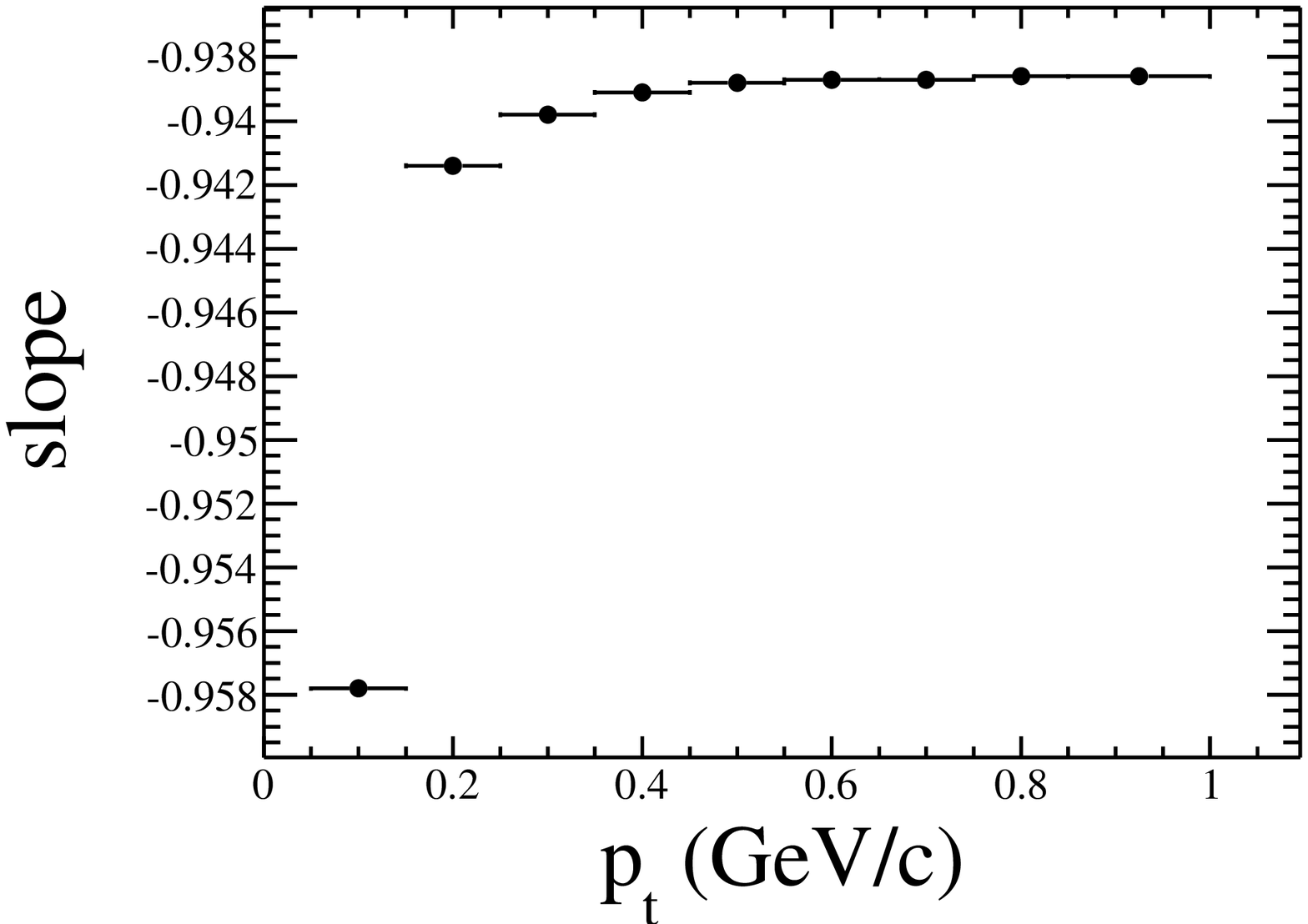}
\figcaption{\label{fig_slope} The slope of the fitted linear line on the $\delta Z$ plane as a function of $p_t$ for the correct three-cluster-combinations from  full simulated $\mu^\pm$ samples.  }
\end{center}

Correct three-cluster-combinations on the $\delta Z$ plane are studied at different transverse momenta with the same simulated $\mu^\pm$ sample.
Figure~\ref{fig5} shows an example for $\mu^\pm$ with $p_t$ between 0.15 and 0.25 GeV$/c$ and a nice anti-diagonal pattern emerges which can also be fitted to a linear line passing the origin point.
As shown in Fig.~\ref{fig_slope}, the slope of the fitted line has a slight transverse momentum $p_t$ dependence.
Similarly, the signed distance $d_{\delta Z}$ to the linear line and projected position $\delta L_{\delta Z}$ for three-cluster-combinations on the $\delta Z$ plane are also defined as shown in Fig.~\ref{fig_d_dL}, and the mean and resolution of $d_{\delta Z}$ distributions at different $\delta L_{\delta Z}$ and $p_t$ are studied by fits to a Gaussian function (one example is shown in Fig. 8).
The obtained mean values of $d_{\delta Z}$ on the $\delta Z$ plane are consistent with zero.
The resolution of $d_{\delta Z}$ on the $\delta Z$ plane for a certain $p_t$ can be fitted to a second order polynomial function with even terms only as shown in Fig.~\ref{fig8}.
\begin{center}
\includegraphics[width=0.4\textwidth]{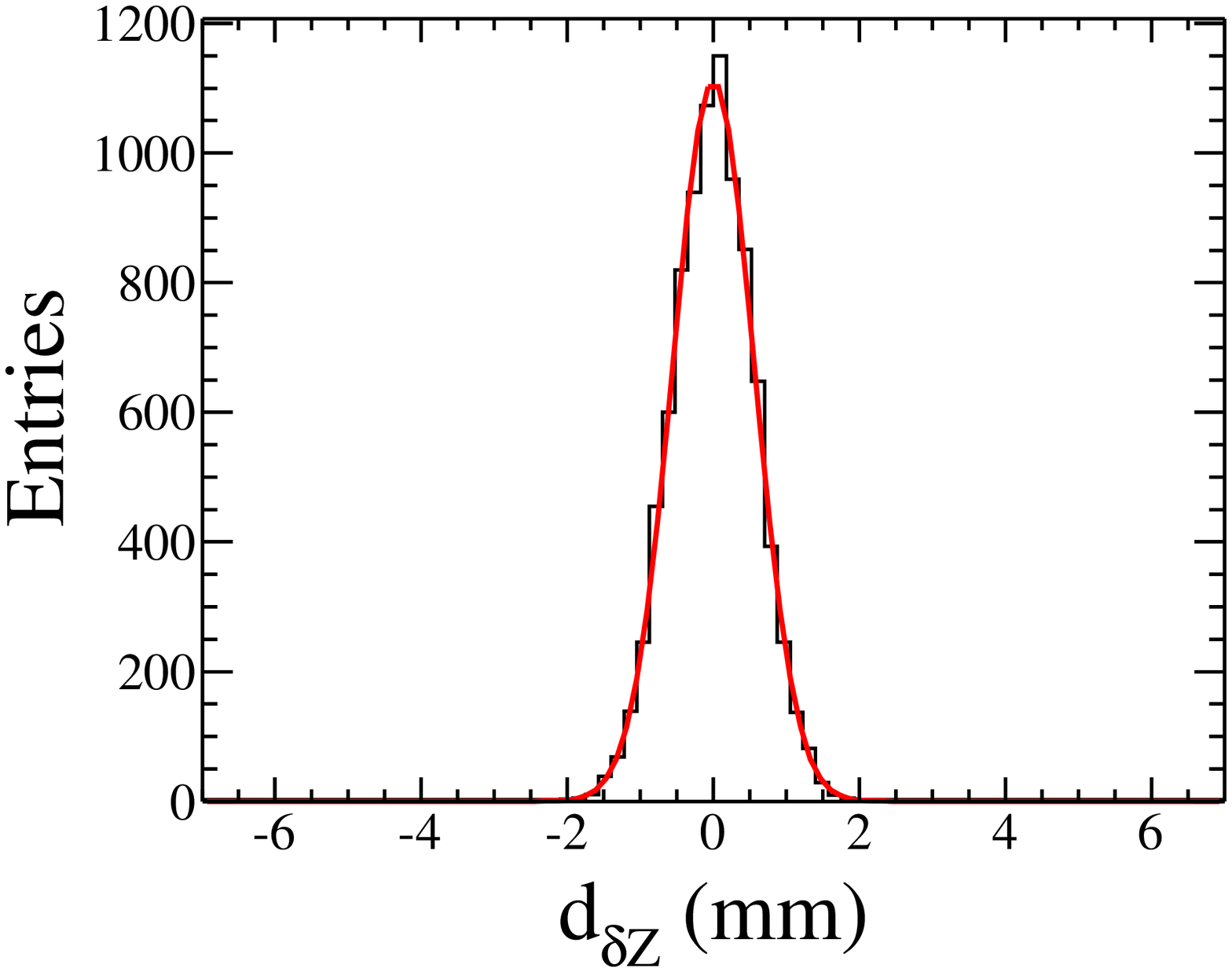}
\figcaption{\label{fig_dz_dL} The $d_{\delta Z}$ distribution for $0<\delta L_{\delta Z}<34.3$ mm. The curve is the fitted result to a Gaussian function.}
\end{center}
\begin{center}
\includegraphics[width=0.4\textwidth]{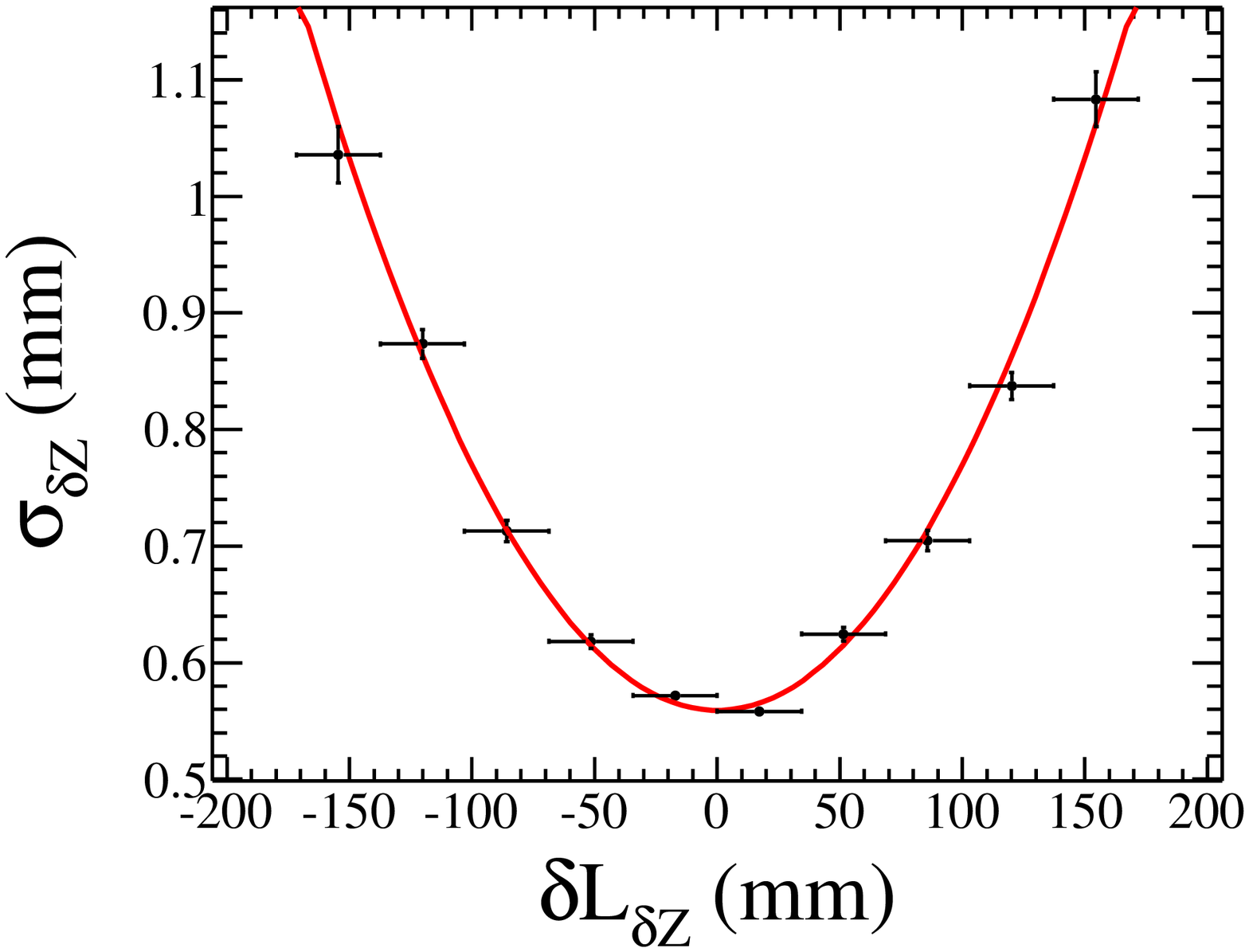}
\figcaption{\label{fig8} The resolution $\sigma_{\delta Z}$ of $d_{\delta Z}$ at different $\delta L_{\delta Z}$ for $\mu^\pm$ with $p_t$ between 0.15 and 0.25~GeV$/c$ (black dots with error bar). The curve is the fitted result to a second order polynomial with even terms only.}
\end{center}

For $\mu^\pm$ with a certain $p_t$ interval, the $\pm 3\sigma_{\delta Z}$ area for three-cluster-combination selection can be determined with the corresponding slope and resolution. But before the track reconstruction, the momentum of charged particles is unknown.
So the $3\sigma_{\delta Z}$ areas for different $p_t$ are merged together to define a selection window for charged tracks with $p_t>0.05$~GeV$/c$, which is shown in Fig.~\ref{fig5}.

\subsection{Efficiency of track segment finding}

With the defined selection windows, as described in the previous subsections, on both of the $\delta\phi$ and $\delta Z$ planes,
the three-cluster-combinations are selected as the track segment candidates in CGEM-IT. All the three-cluster-combinations
and those after the selection are shown in Fig.~\ref{delPhiZ_sel} for an independent full simulated $\mu^\pm$ samples.
The three-cluster-combinations inconsistent with the found patterns are due to the clusters induced by the electrons from the muon decay.
The efficiency of the track segment finding is computed as the ratio of the number of the correct-three-cluster-combinations after selection over the number of generated tracks. The obtained efficiency of the track segment finding as a function of $p_t$ for simulated $\mu^\pm$ is shown in Fig.~\ref{fig9}, which ranges from $99.1\%$ to $99.7\%$ depending on $p_t$. The lower efficiency at lower $p_t$ is caused by the decay of muon.
\begin{center}
\includegraphics[width=0.23\textwidth]{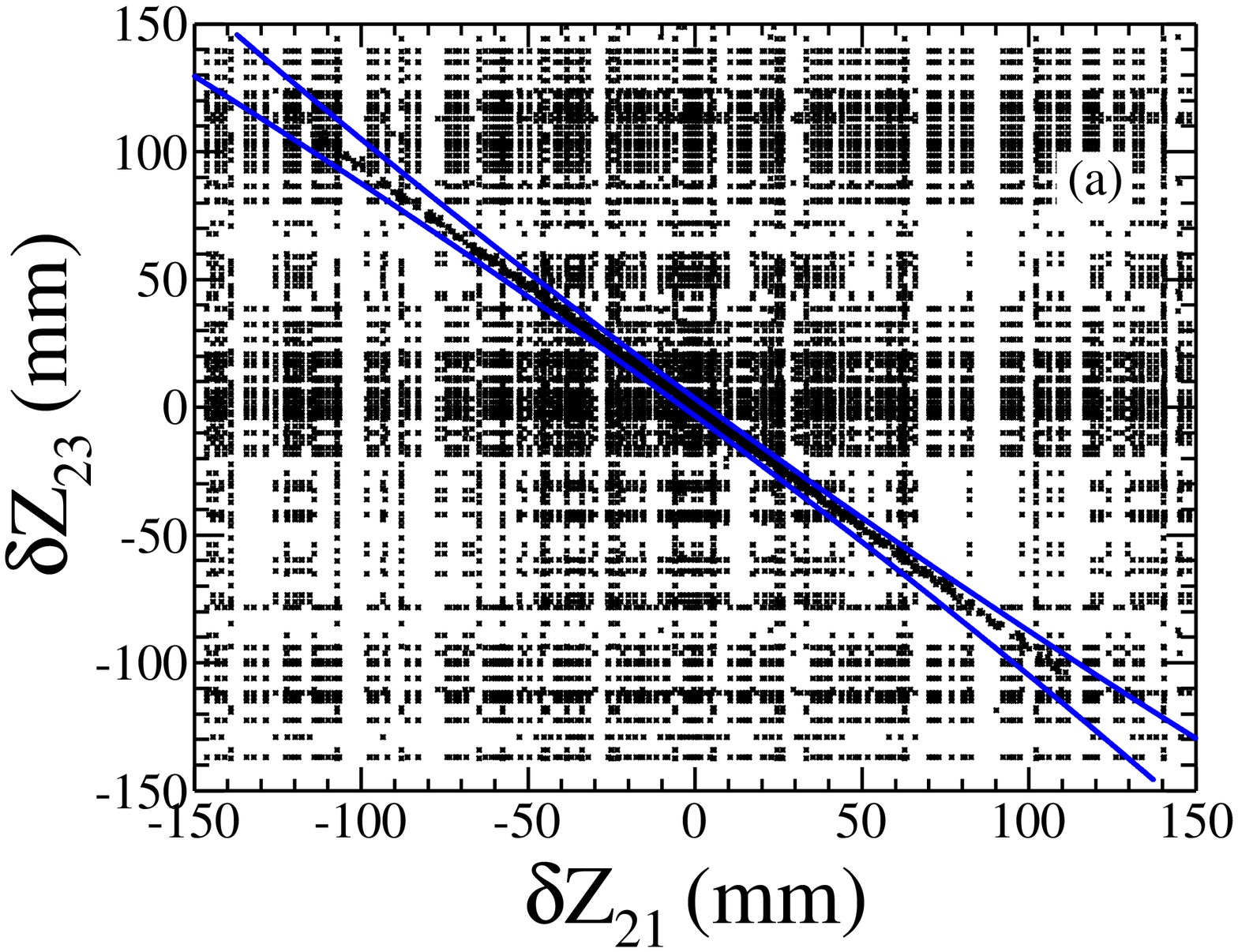}
\includegraphics[width=0.23\textwidth]{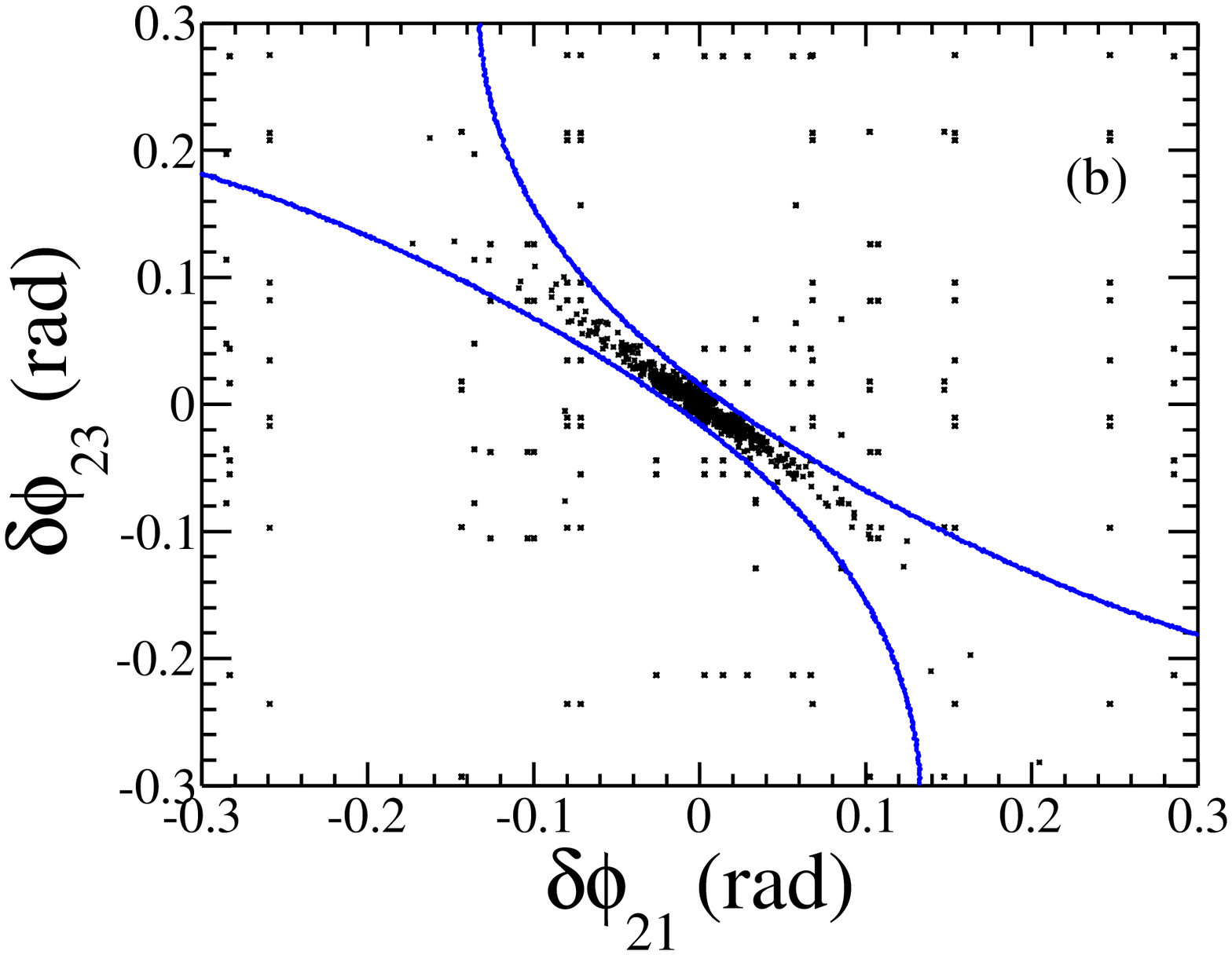}
\includegraphics[width=0.23\textwidth]{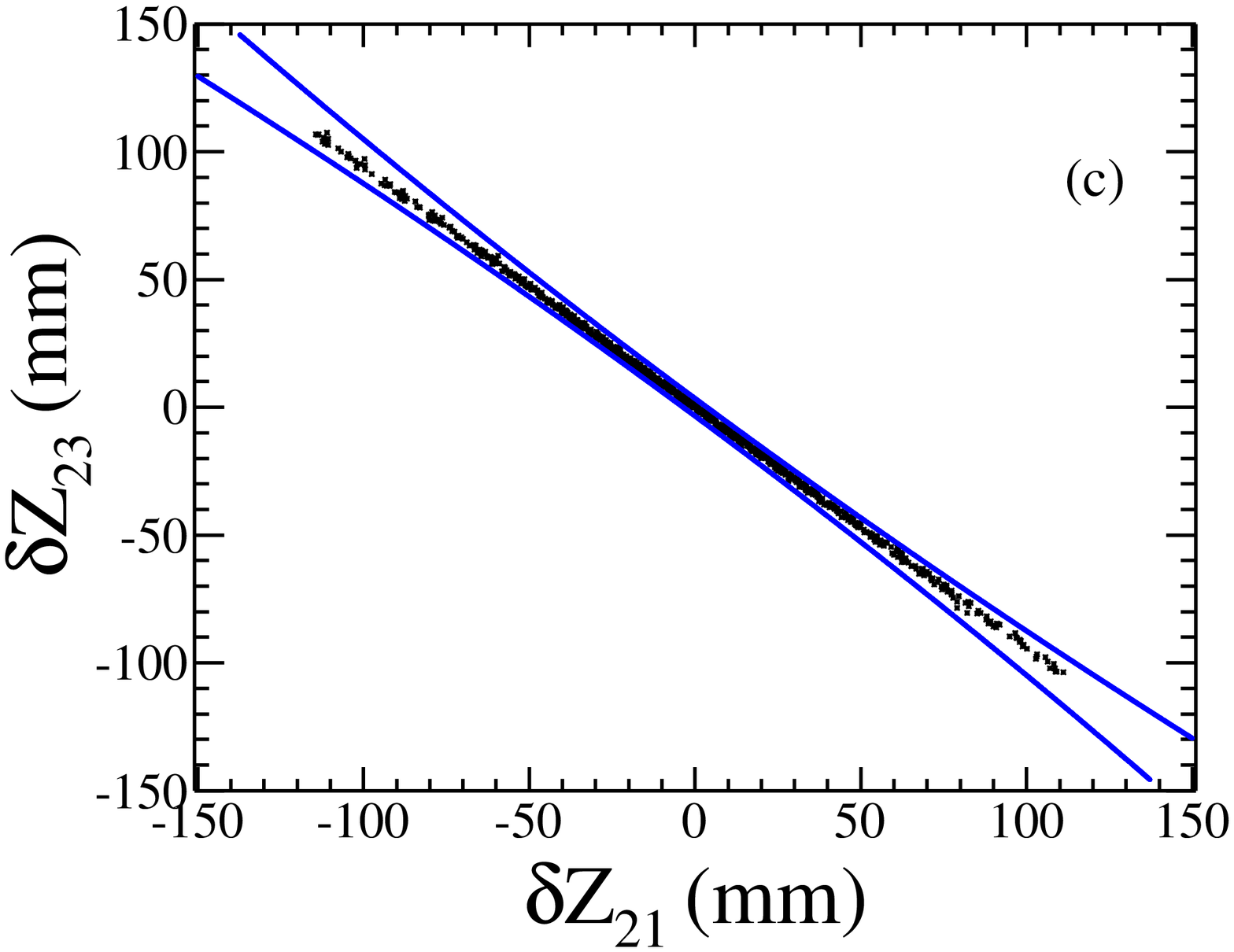}
\includegraphics[width=0.23\textwidth]{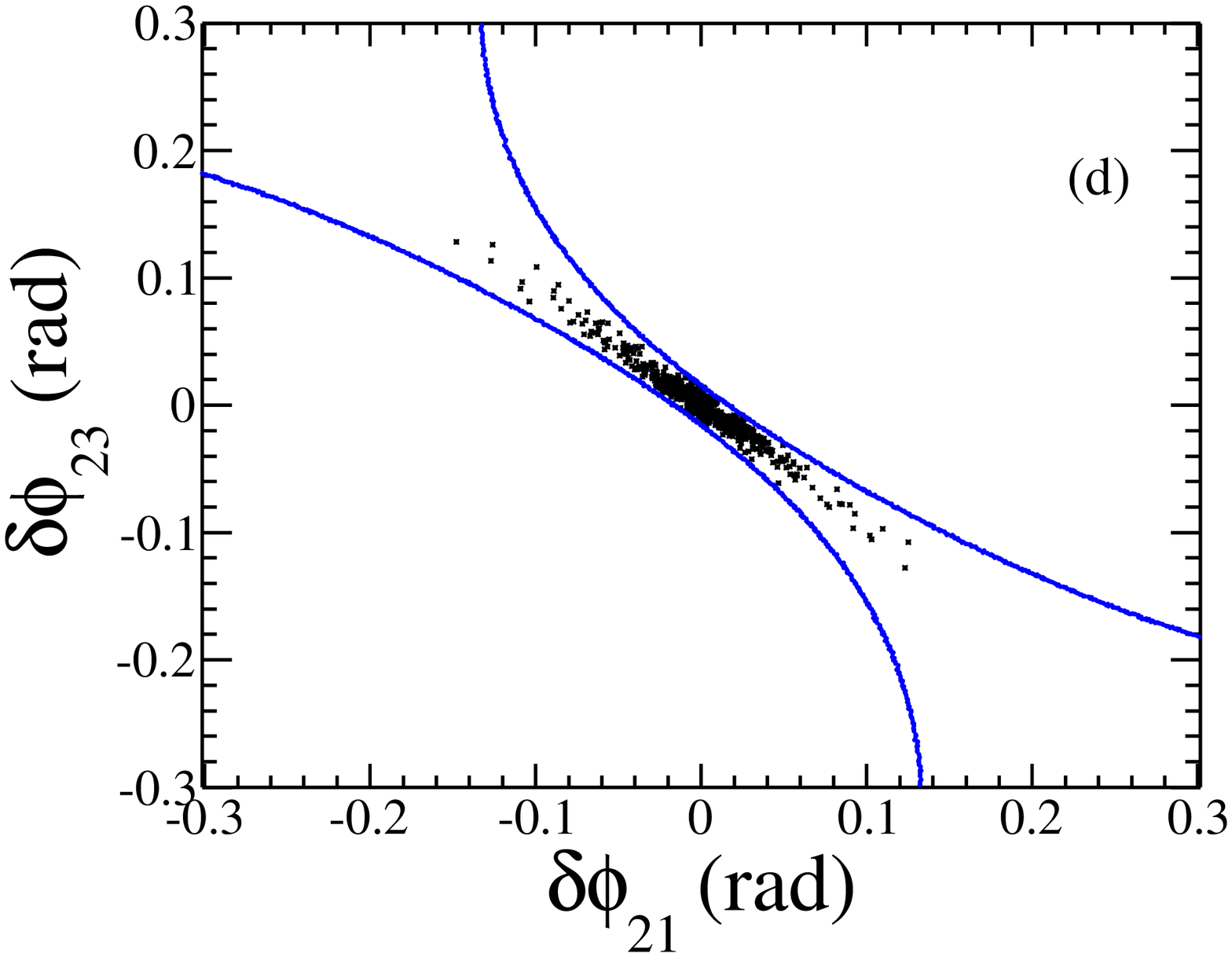}
\figcaption{\label{delPhiZ_sel}(color online) The distributions of three-cluster-combinations for $\mu^{\pm}$ with $p_{t}>0.05~$GeV$/c$. Before the selection, three-cluster-combinations are showed on the $\delta Z$ (a) and $\delta\phi$ (b) plane. Most correct three-cluster-combinations are selected by both selection windows on the $\delta Z$ and $\delta\phi$ plane, and their distribution are (c) and (d). The backgrounds are incorrect three-cluster-combinations in which some clusters are produced by the electrons from the muon decay.}
\end{center}
\begin{center}
\includegraphics[width=0.4\textwidth]{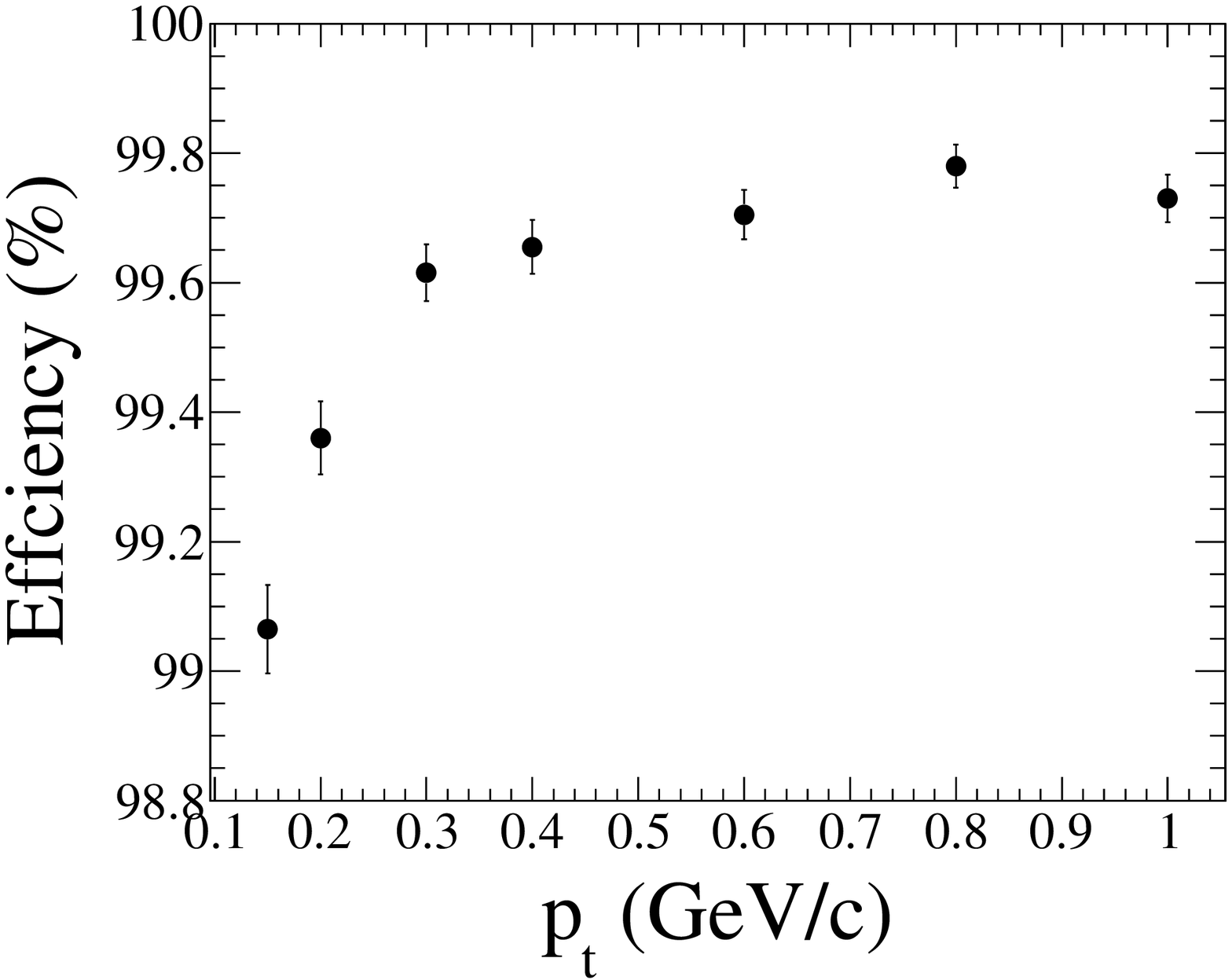}
\figcaption{\label{fig9} The efficiency of the track segment finding as a function of $p_t$ for full simulated $\mu^\pm$ samples.}
\end{center}

\subsection{Track segment fitting}

To describe quantitatively the found track segment in CGEM-IT,
the three-cluster-combinations are fitted to a helix model with five parameters.
Some of the parameters are defined with respect to a referenced three-dimensional point called pivot, which can be
arbitrary but usually is the interaction point or the origin point.
The five parameters are\\
\indent$dr$: the signed distance on the transverse plane from the pivot to the helix,\\
\indent$\phi_{0}$: the azimuthal angle specifying the pivot with respect to the helix center,\\
\indent$\kappa$: the product of the particle charge and the reciprocal of $p_{t}$,\\
\indent$dz$: the signed distance of the helix from the pivot in the $Z$ direction,\\
\indent$\tan\lambda$: the slope of the track and $\lambda=\pi/2-\theta$ where $\theta$ is the polar angle of the track.\\

The Least Square Method is used for the helix fitting, and the $\chi^{2}$ for the fitting is defined as
\begin{equation}
	\chi^{2}_{fit} = \Sigma_{j}\Sigma_{i}\frac{(X_{i,j}(H)-X_{i,j,measured})^{2}}{\sigma_{i,j}^{2}} \label{a5}
\end{equation}
where $X(H)$ is the cluster position calculated with the helix parameters $H$, $X_{measured}$ means the cluster position measured by the CGEM-IT, $\sigma$ which comes from the residual between the measured result and MC truth information is the error of the measured cluster positions, $i=1,~2$ stand for $\phi$ and $Z$, and $j=1,2,3$ stand for 3 clusters. The $\chi^2_{fit}$ minimization is realized with MINUIT~\cite{minuit}.

\begin{center}
\includegraphics[width=0.4\textwidth]{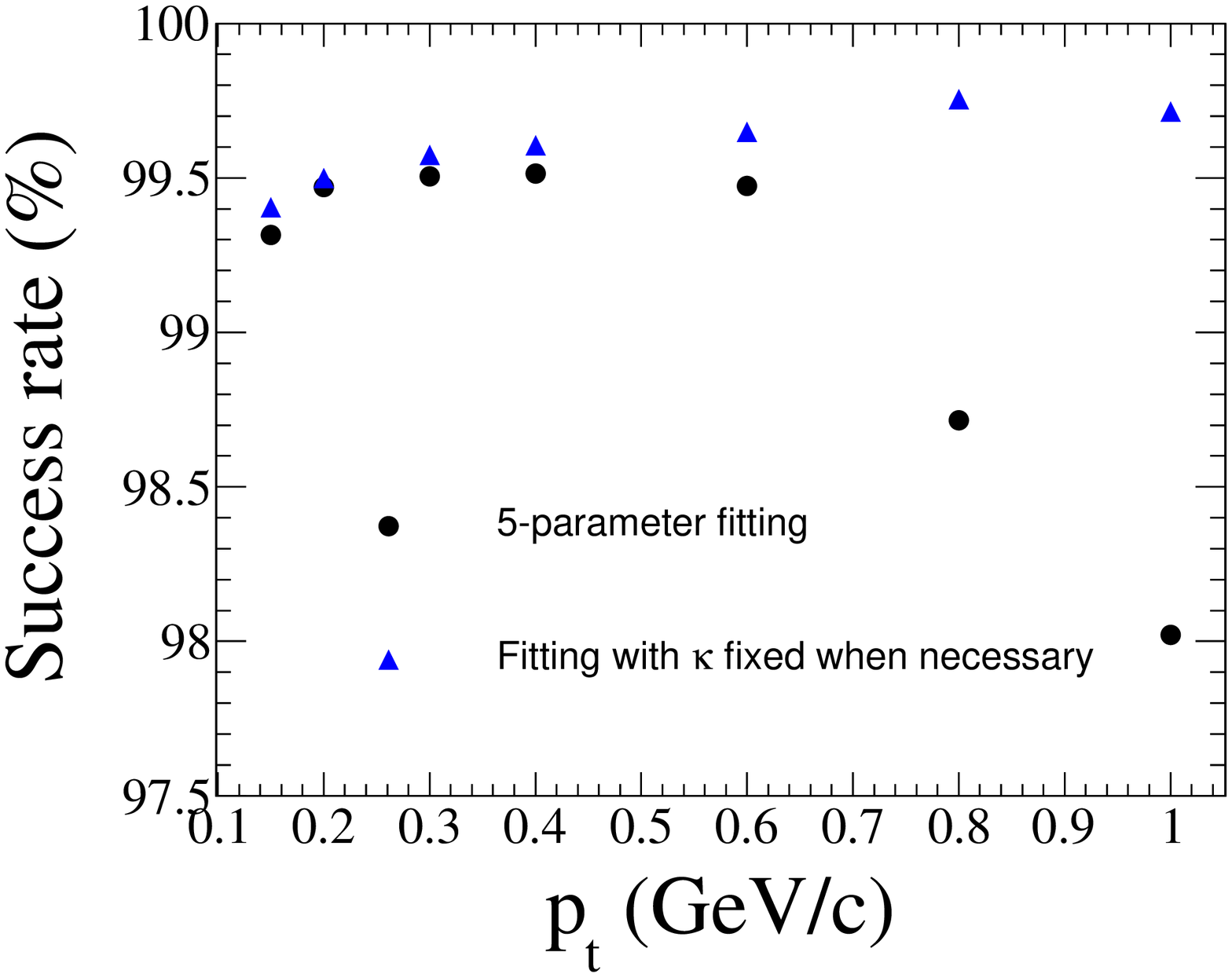}
\figcaption{\label{fig_fitseg} (color online) The success rate of the helix fitting as a function of $p_{t}$ with all 5 parameters floating (black dots) and with $\kappa$ fixed if the previous 5-parameter fitting fails (blue triangles).}
\end{center}

Since the high $p_{t}$ track is quite straight in CGEM-IT, only three clusters close-by with limited spatial resolution can
not determine $\kappa$ well. The effect is reflected on the success rate of the 5-parameter helix fitting which drops at high $p_t$ as shown in Fig.~\ref{fig_fitseg} (black dots). To improve the success rate of the helix fitting for high $p_t$ tracks,
the $\kappa$ can be fixed to the one calculated with the three transverse positions of the clusters if the 5-parameter helix fitting fails.
The success rate of the helix fitting after fixing $\kappa$ is kept above $99.4\%$ and reaches $99.7\%$ at high $p_t$ as shown in Fig.~\ref{fig_fitseg} (blue triangles).

The residual distributions for the five helix parameters after the track segment fitting are shown in Fig.~\ref{fig_residual},
where the residual is defined by $\delta H_i=H_i^{fit}-H_i^{truth}~(i=1\sim5)$, namely the difference between the fitted helix parameters $H_i^{fit}$ and the true helix parameters $H_i^{truth}$ at the $\mu^\pm$ track generation. These distributions demonstrate that reasonably unbiased track parameters can be obtained by the fitting.

\begin{center}
\includegraphics[width=0.23\textwidth]{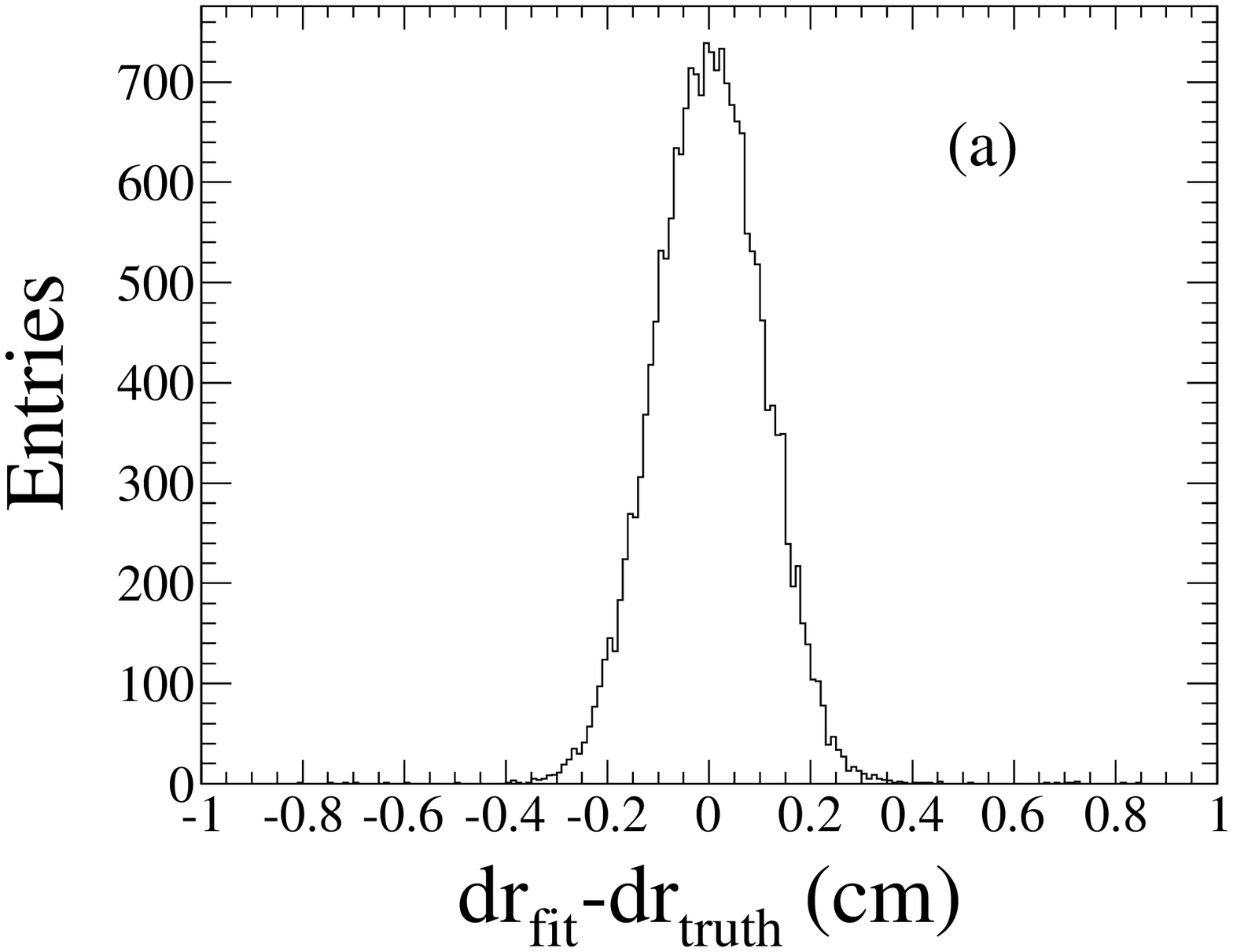}
\includegraphics[width=0.23\textwidth]{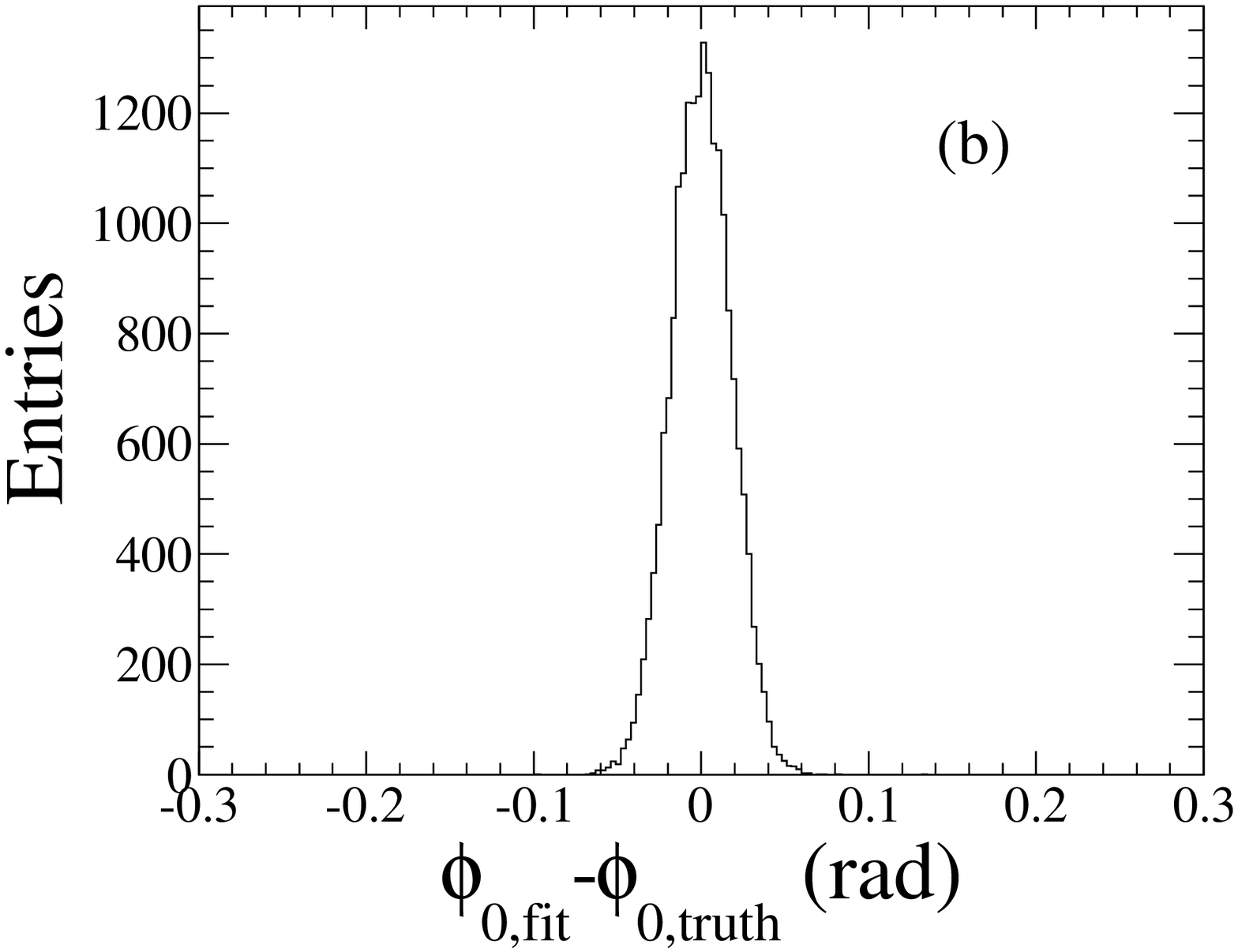}
\includegraphics[width=0.23\textwidth]{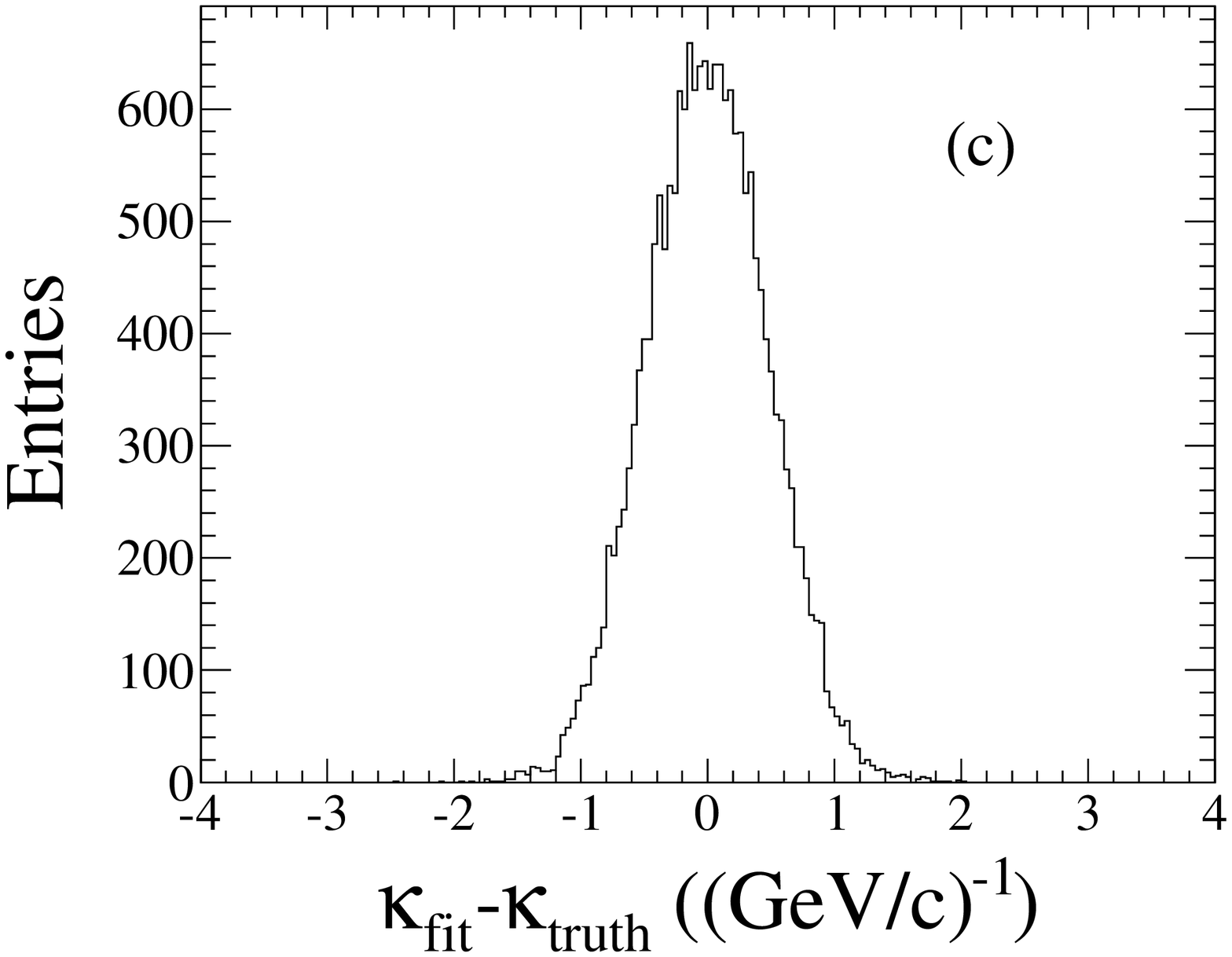}
\includegraphics[width=0.23\textwidth]{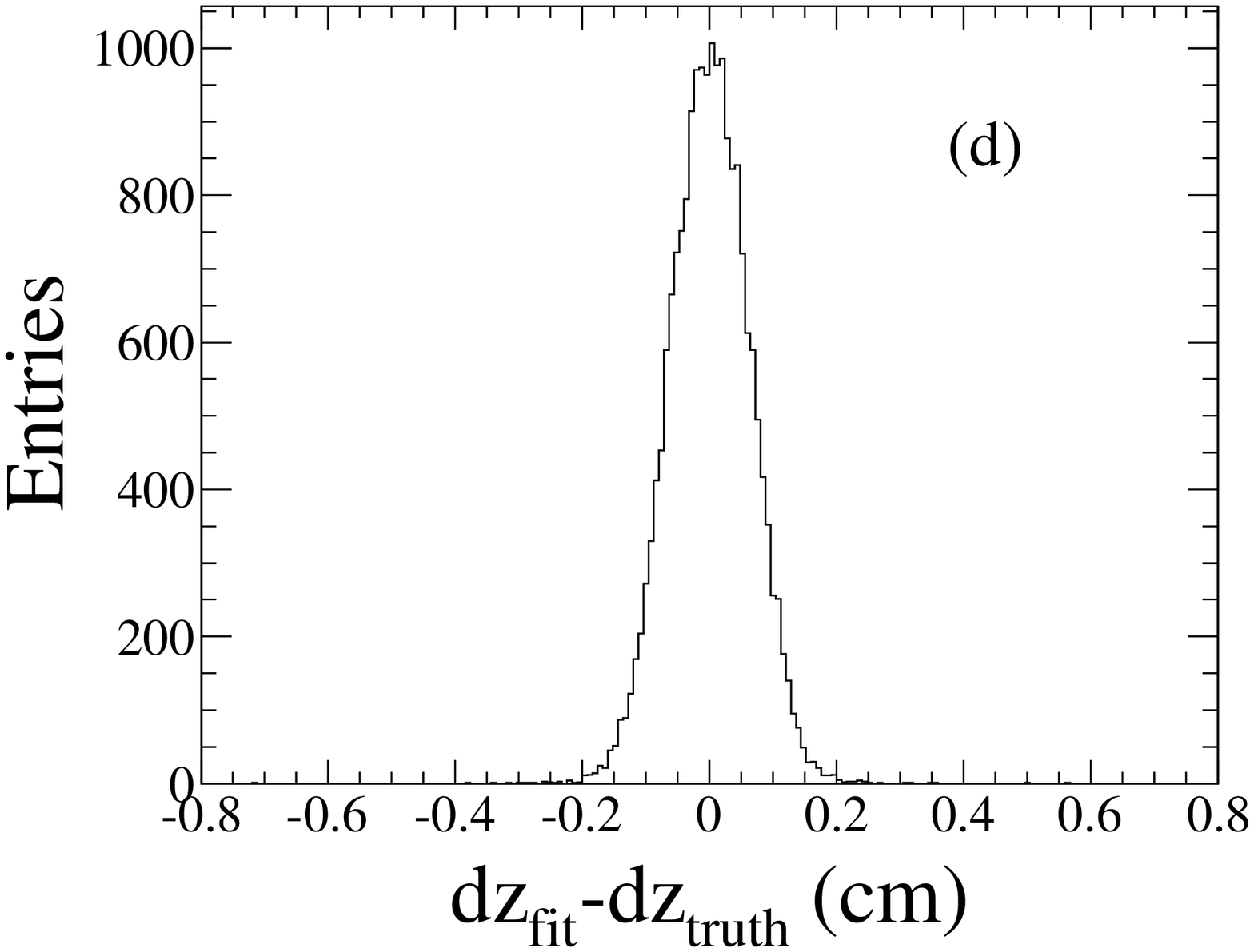}
\includegraphics[width=0.23\textwidth]{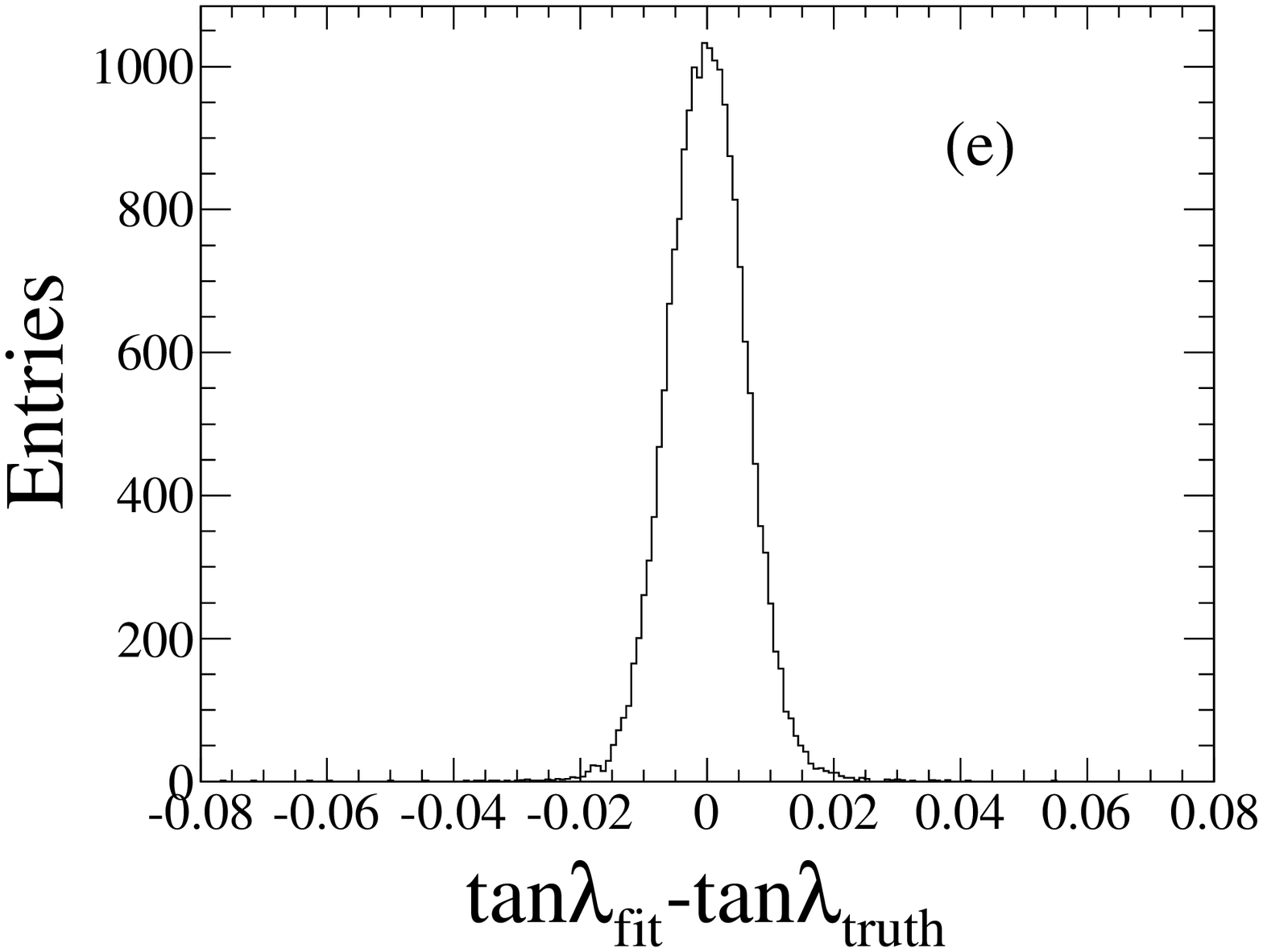}
\figcaption{\label{fig_residual} The distributions of residual for the helix parameter $dr$ (a), $\phi_{0}$ (b), $\kappa$ (c), $dz$ (d) and $tan\lambda$ (e), which are obtained from a full simulated $\mu^\pm$ with $p_t=0.8$~GeV$/c$.}
\end{center}

\section{Matching CGEM-IT track segments to ODC tracks}

Typically, a complete track is found if both of the track segment in CGEM-IT and the track in ODC are found and matched.
The tracks in ODC can be reconstructed by the track finding algorithm~\cite{MDC_track,MDC_track_zy},
the helix parameters of which are also obtained by fitting.
\end{multicols}
\begin{center}
\includegraphics[width=0.3\textwidth]{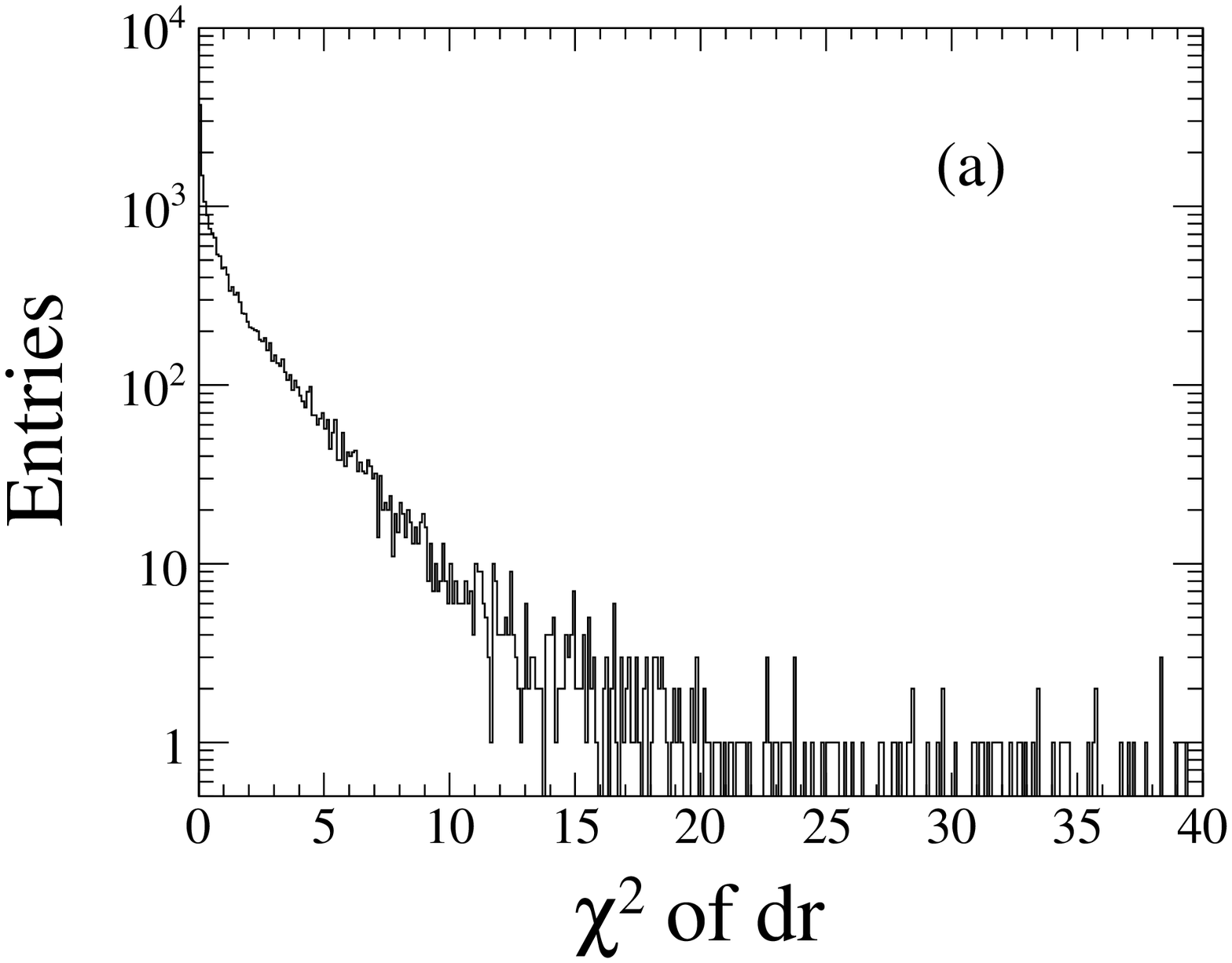}
\includegraphics[width=0.3\textwidth]{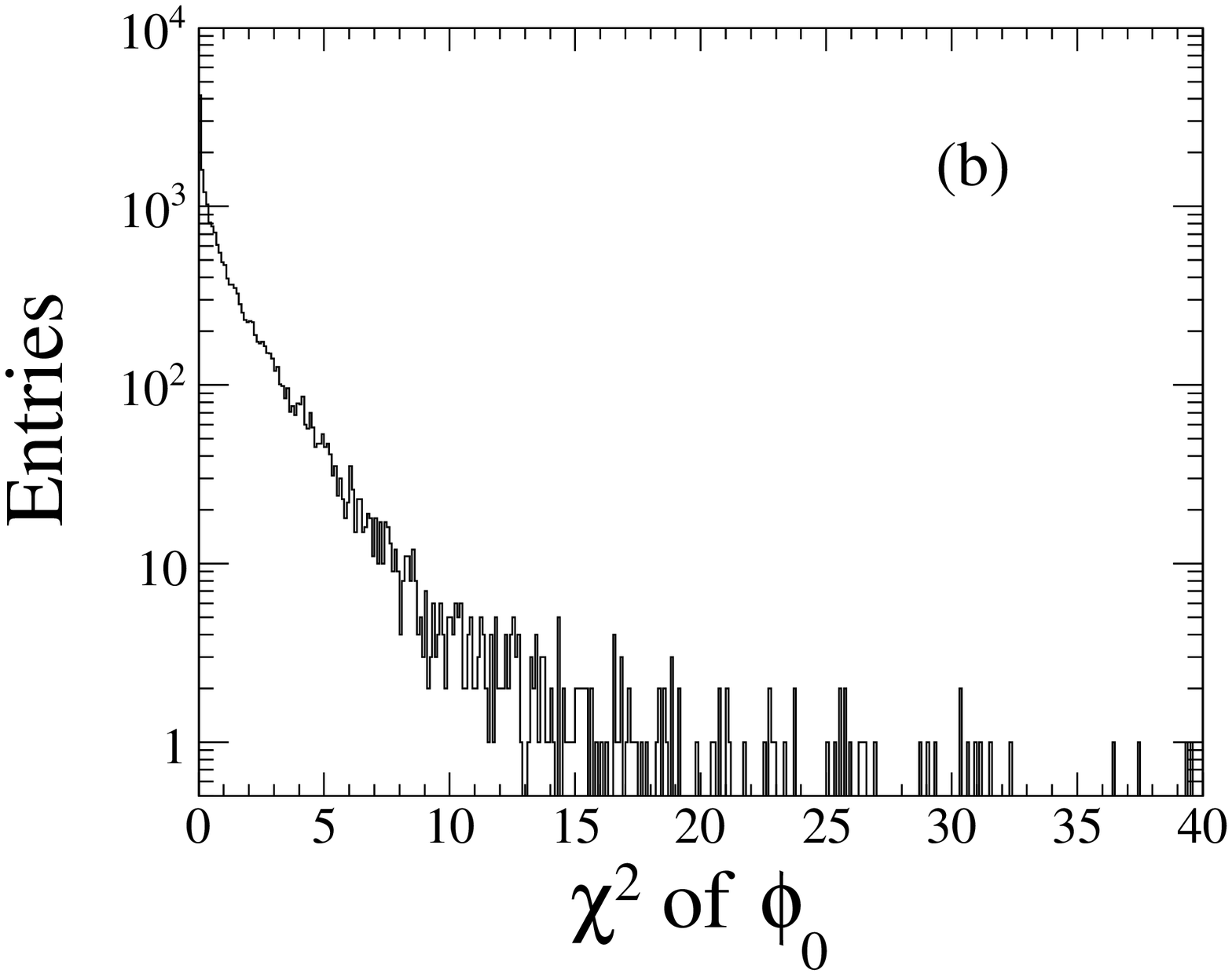}
\includegraphics[width=0.3\textwidth]{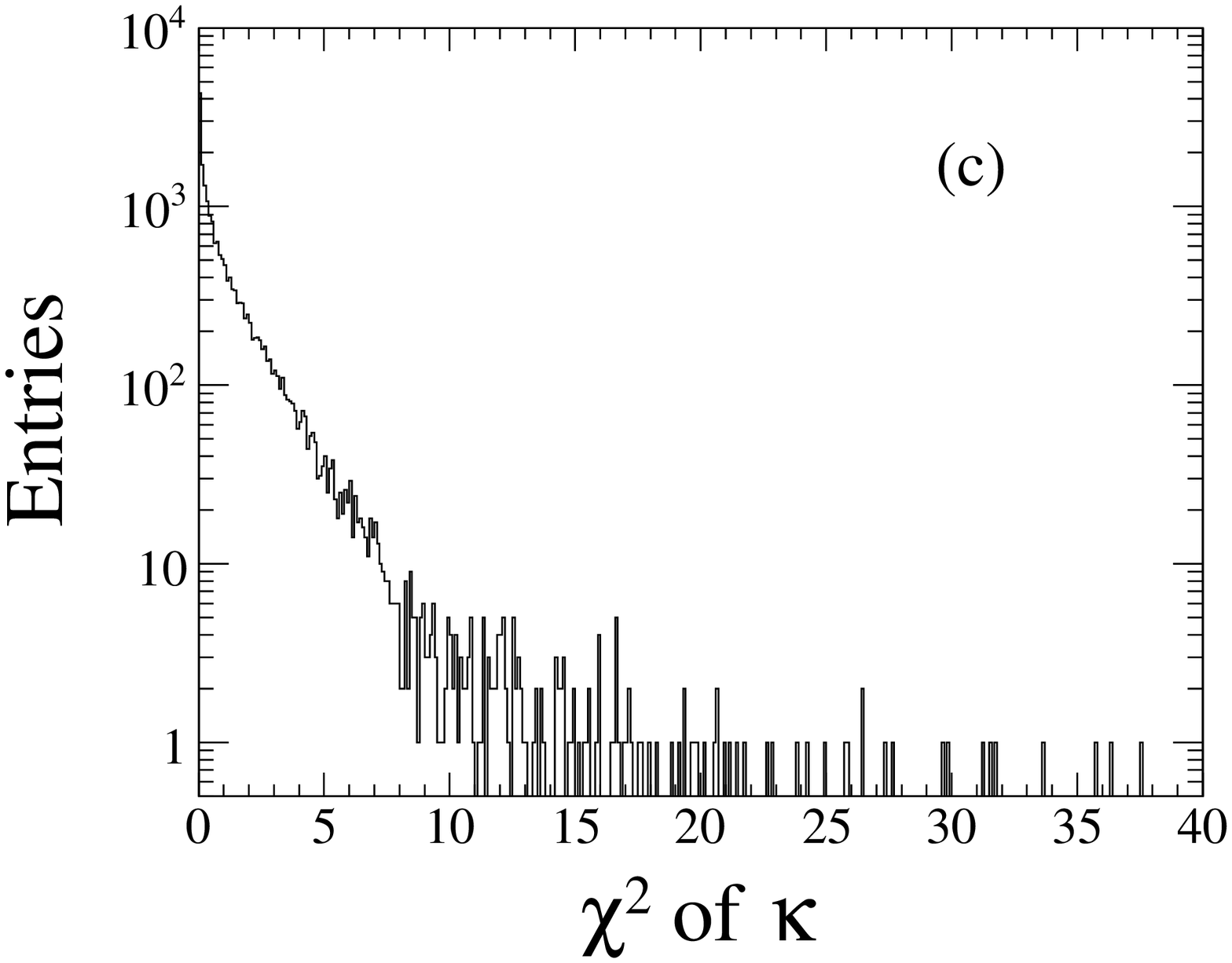}
\includegraphics[width=0.3\textwidth]{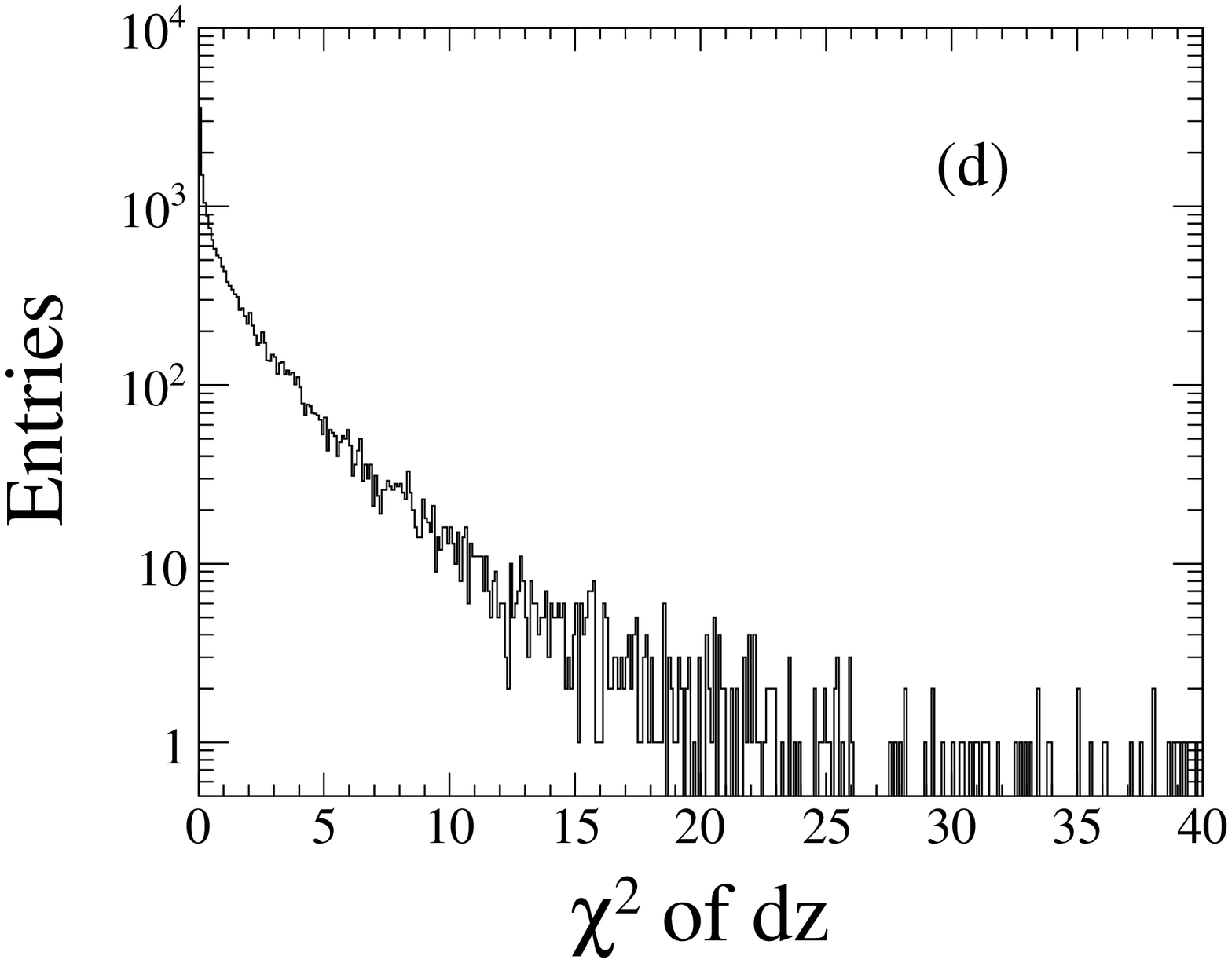}
\includegraphics[width=0.3\textwidth]{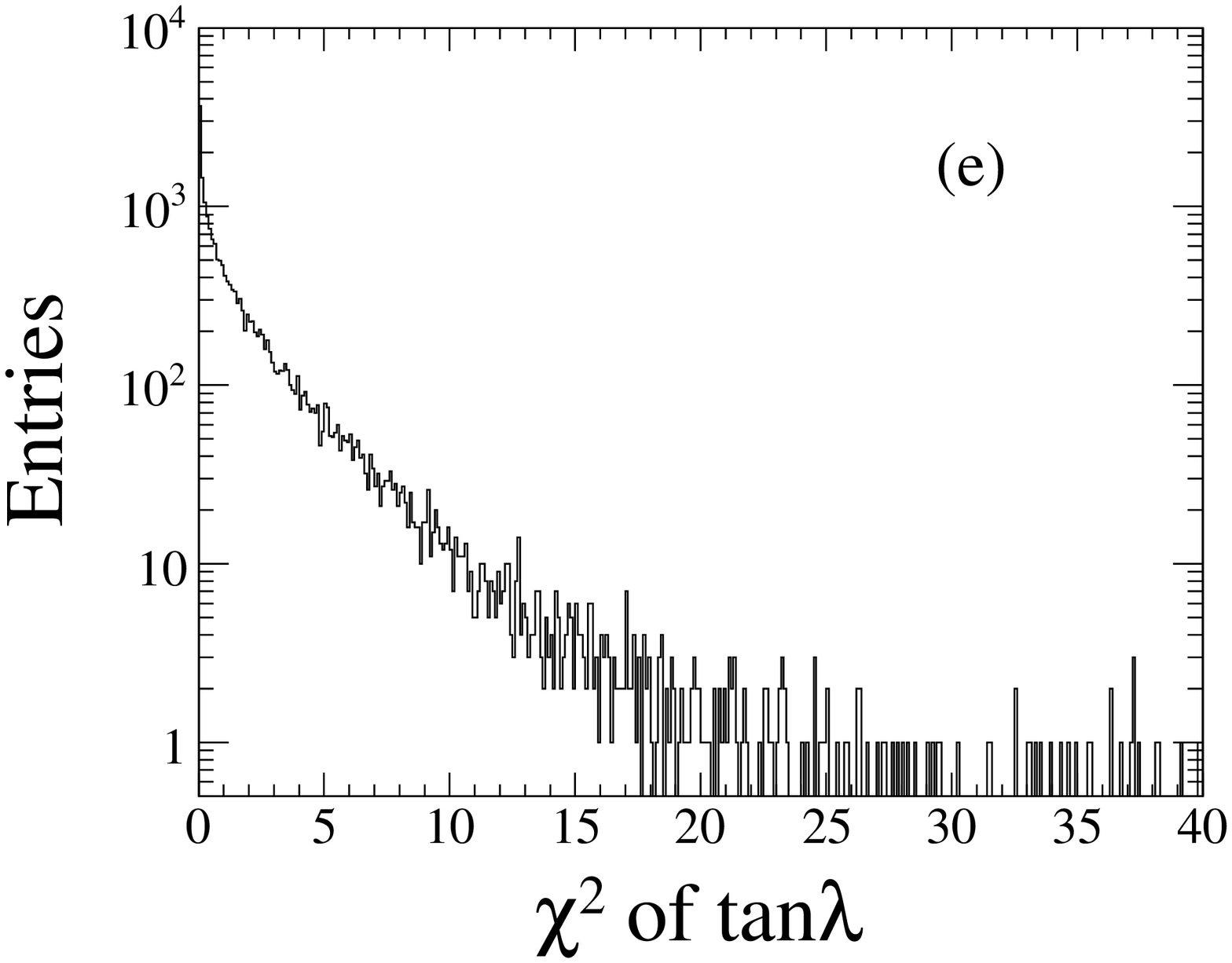}
\figcaption{\label{fig13} The distributions of $\chi^{2}_{match}$ for the helix parameter $dr$ (a), $\phi_{0}$ (b), $\kappa$ (c), $dz$ (d) and $tan\lambda$ (e), which are obtained from a full simulated $\mu^\pm$ sample with $p_t=0.8$~GeV$/c$.}
\end{center}
\begin{center}
\includegraphics[width=0.3\textwidth]{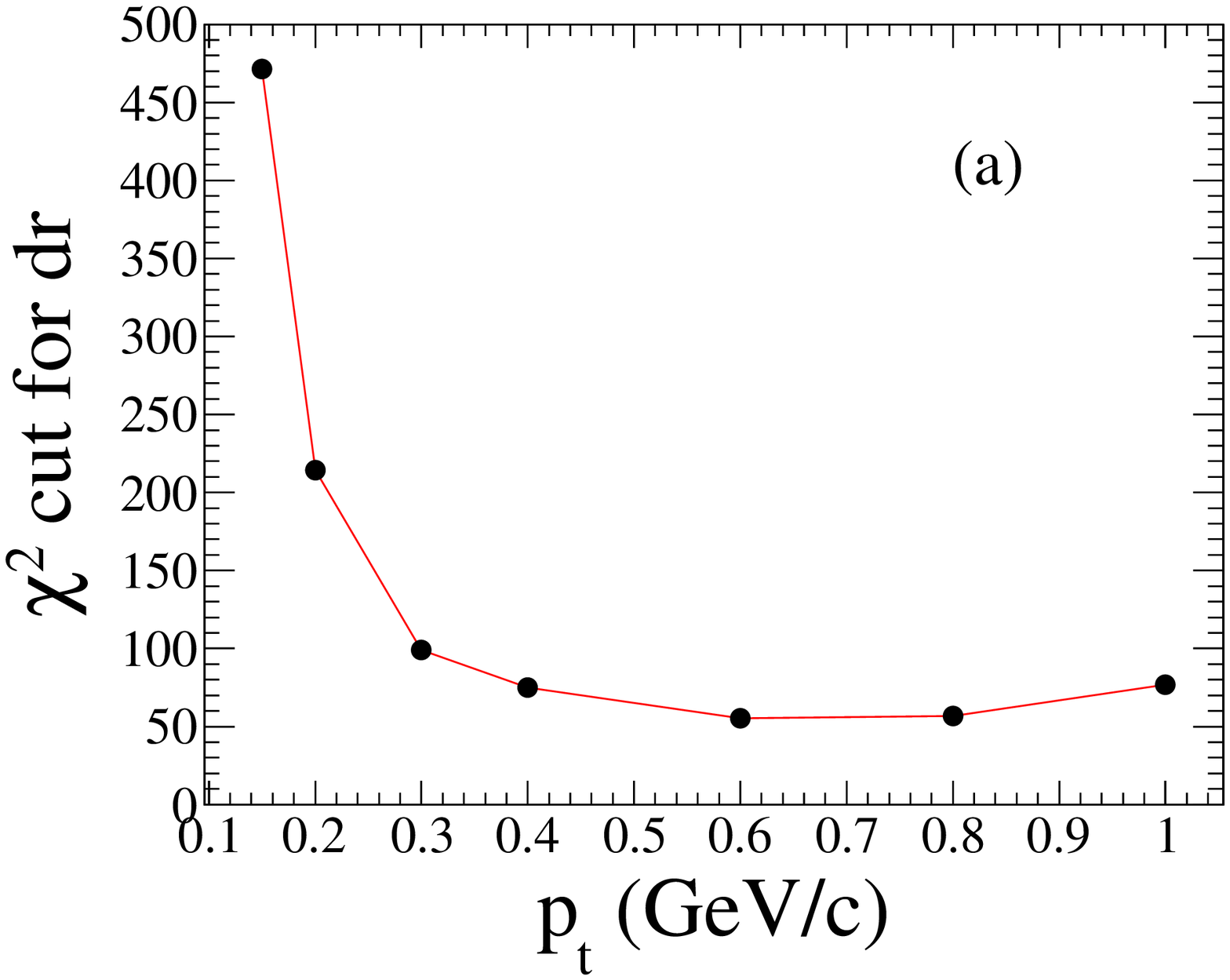}
\includegraphics[width=0.3\textwidth]{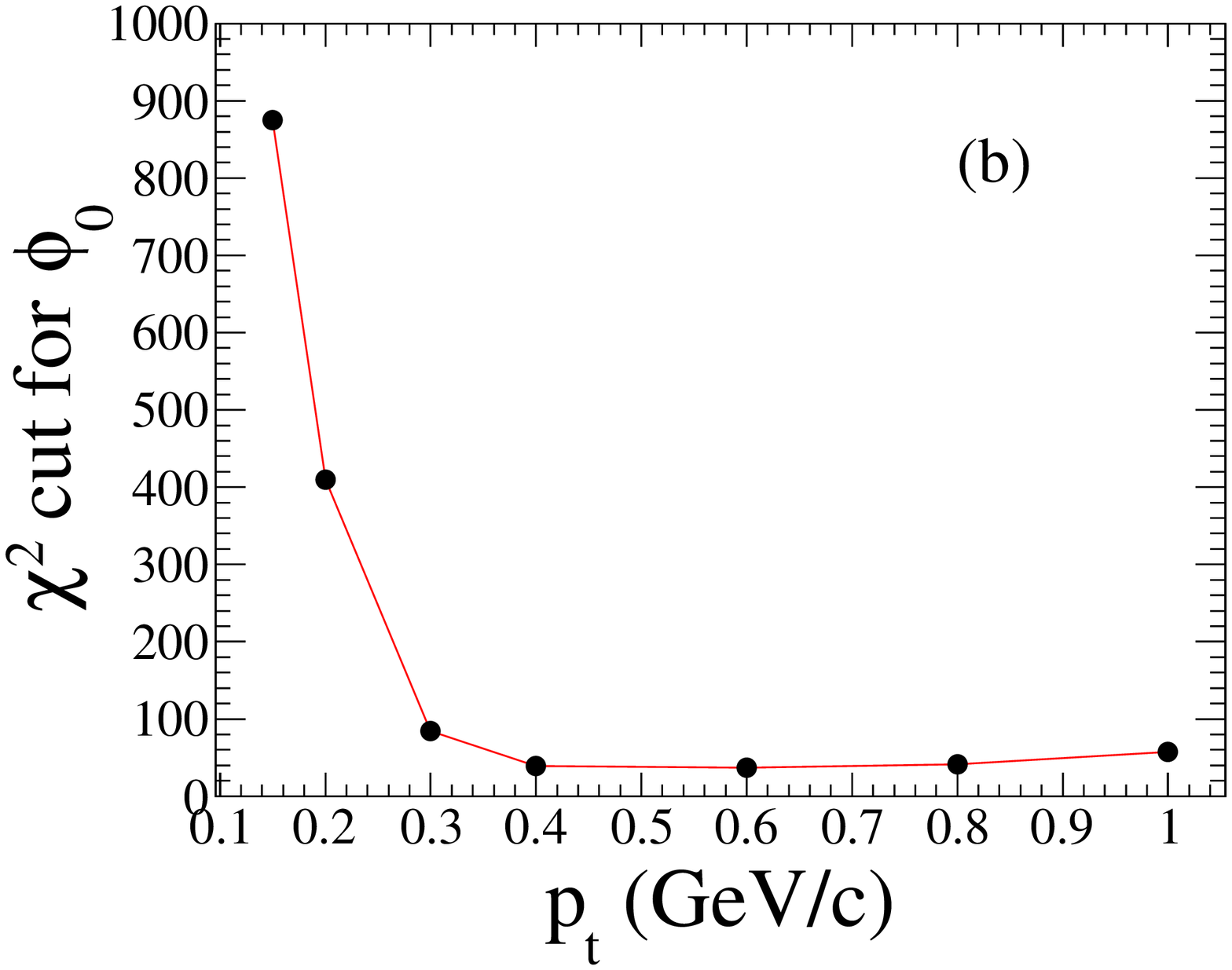}
\includegraphics[width=0.3\textwidth]{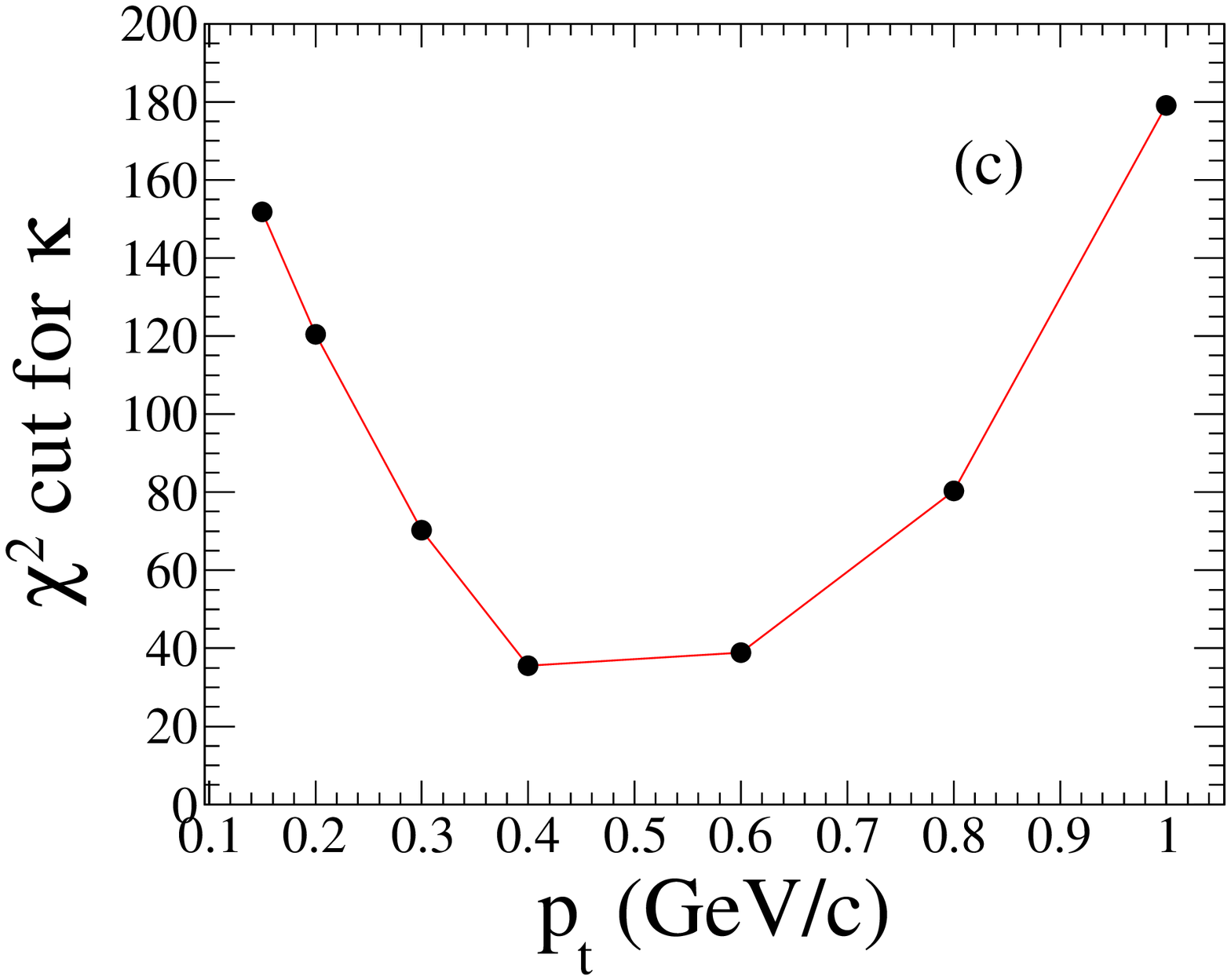}
\includegraphics[width=0.3\textwidth]{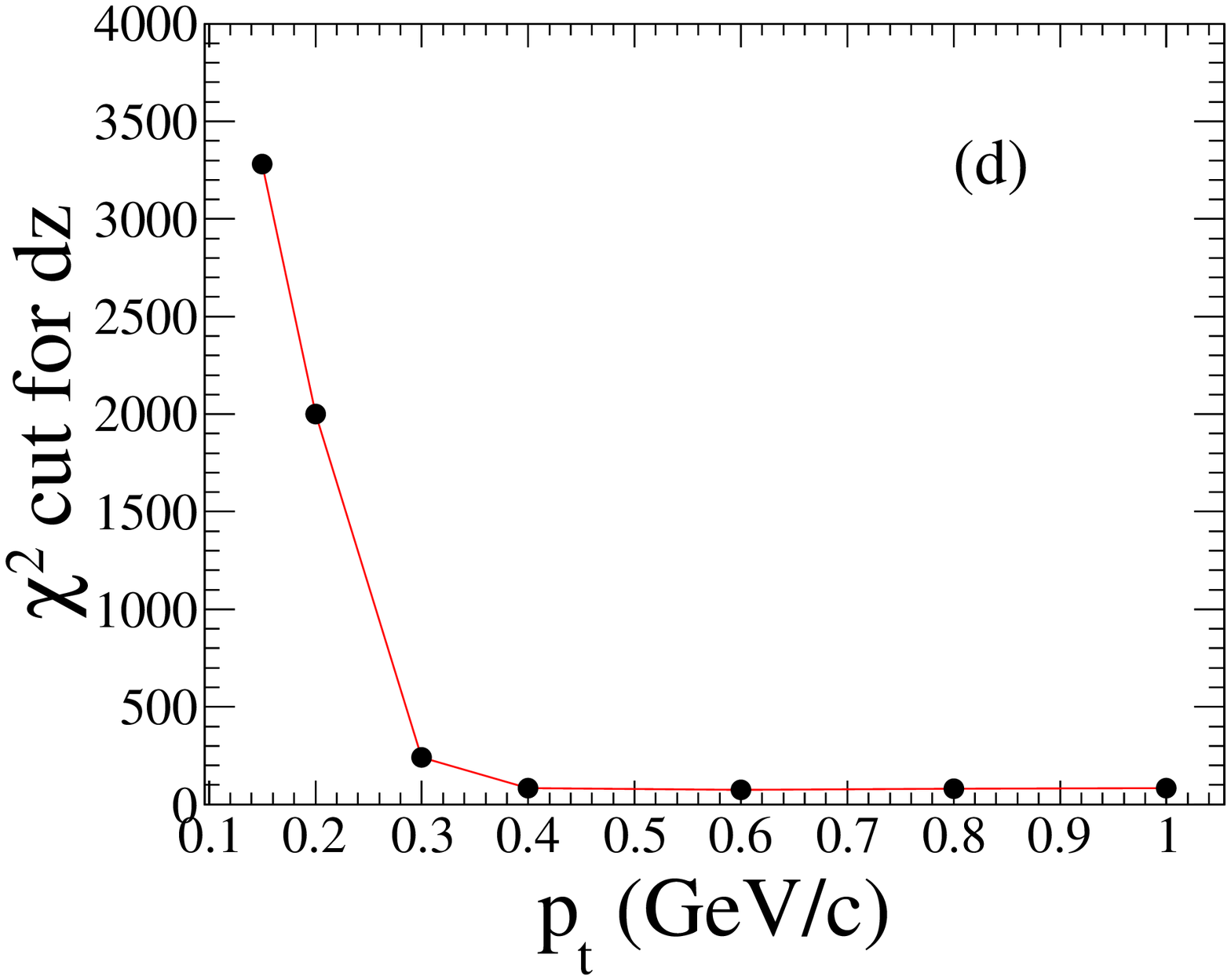}
\includegraphics[width=0.3\textwidth]{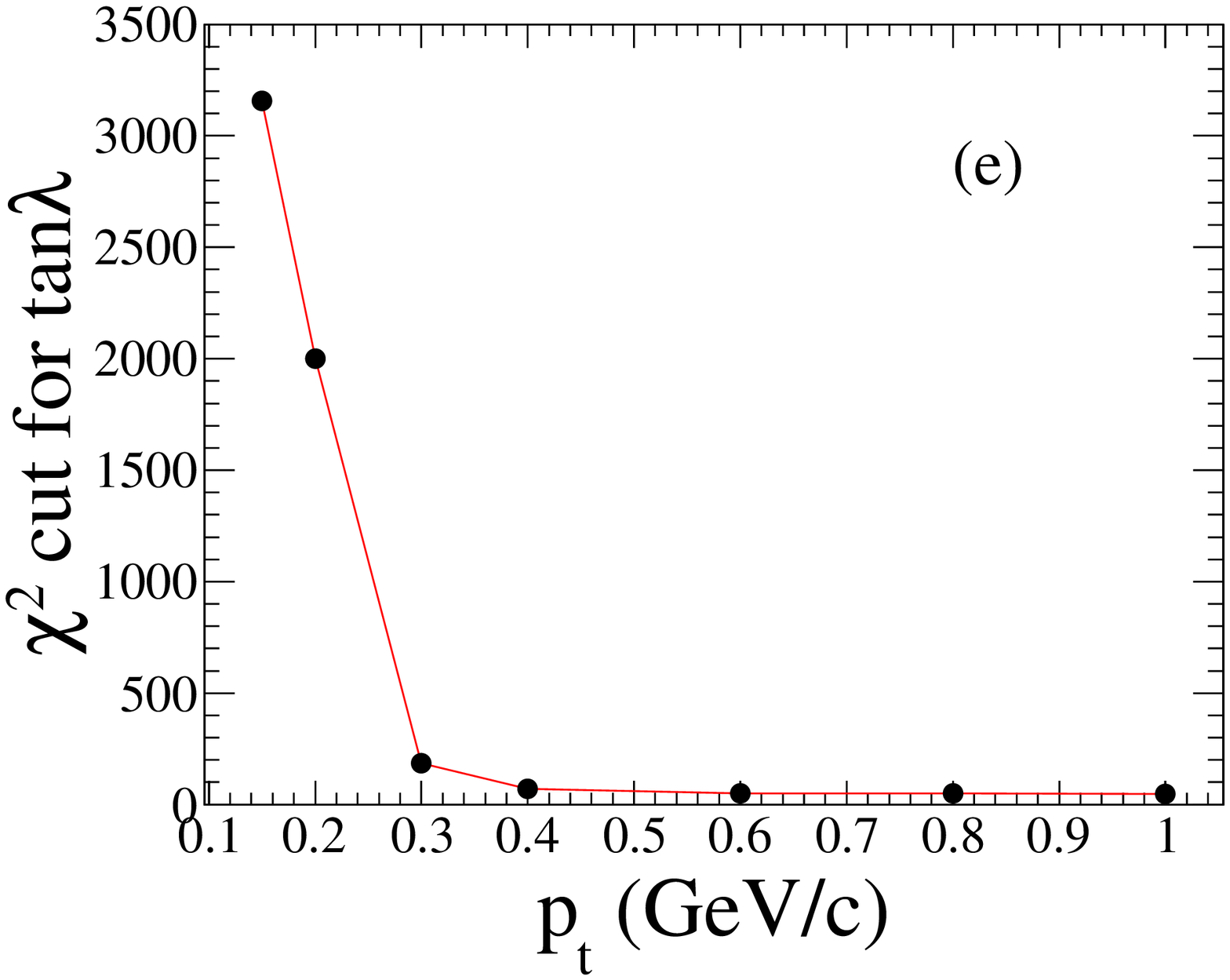}
\figcaption{\label{fig14} The $\chi^2_{cut}$ values at different $p_t$ for $dr$ (a), $\phi_{0}$ (b), $\kappa$ (c), $dz$ (d) and $tan\lambda$ (e).}
\end{center}
\newpage
\begin{multicols}{2}

\noindent To evaluate the consistency between the track segments in
CGEM-IT and the tracks in ODC, quantity $\chi^2_{match}$ is calculated for each of the 5 helix parameters:
\begin{equation}
	\chi^{2}_{match} = \frac{(H_{CGEM-IT}-H_{ODC})^{2}}{\sigma_{CGEM-IT}^2 + \sigma_{ODC}^2} \label{a4}
\end{equation}
where $H$ is a parameter of helix from the track segment in CGEM-IT or the track in ODC,
$\sigma$ is the corresponding error.
A global $\chi^2$ of the five track parameters can be defined. But the error matrices are not perfectly evaluated here, as some effects can not be considered in a simple Helix fitting (for instance, the inhomogeneity of magnetic field, energy loss of charged tracks and so on). And this affects the five parameters unequally, so it is better to study the parameters independently.
As an example, the $\chi^2_{match}$ distributions of the 5 parameters for a simulated $\mu^\pm$ sample with $p_t=0.8$~GeV$/c$ are shown in Fig.~\ref{fig13}. Appropriate $\chi^2_{cut}$ values are chosen to ensure that the individual efficiency for each parameter is not less than $99.5\%$ when $\chi^2_{match}<\chi^2_{cut}$ is required. The determined $\chi^2_{cut}$ values at different $p_t$ for the 5 helix parameters are shown in Fig.~\ref{fig14}.

The $\chi^2_{cut}$ value at a $p_t$ between two neighboring points on the plot is obtained by a linear interpolation.
If each of the five $\chi^2_{match}$ are less than the corresponding $\chi^2_{cut}$ value, the track segment in the CGEM-IT
and the track in ODC are called matched.
\begin{center}
\includegraphics[width=0.4\textwidth]{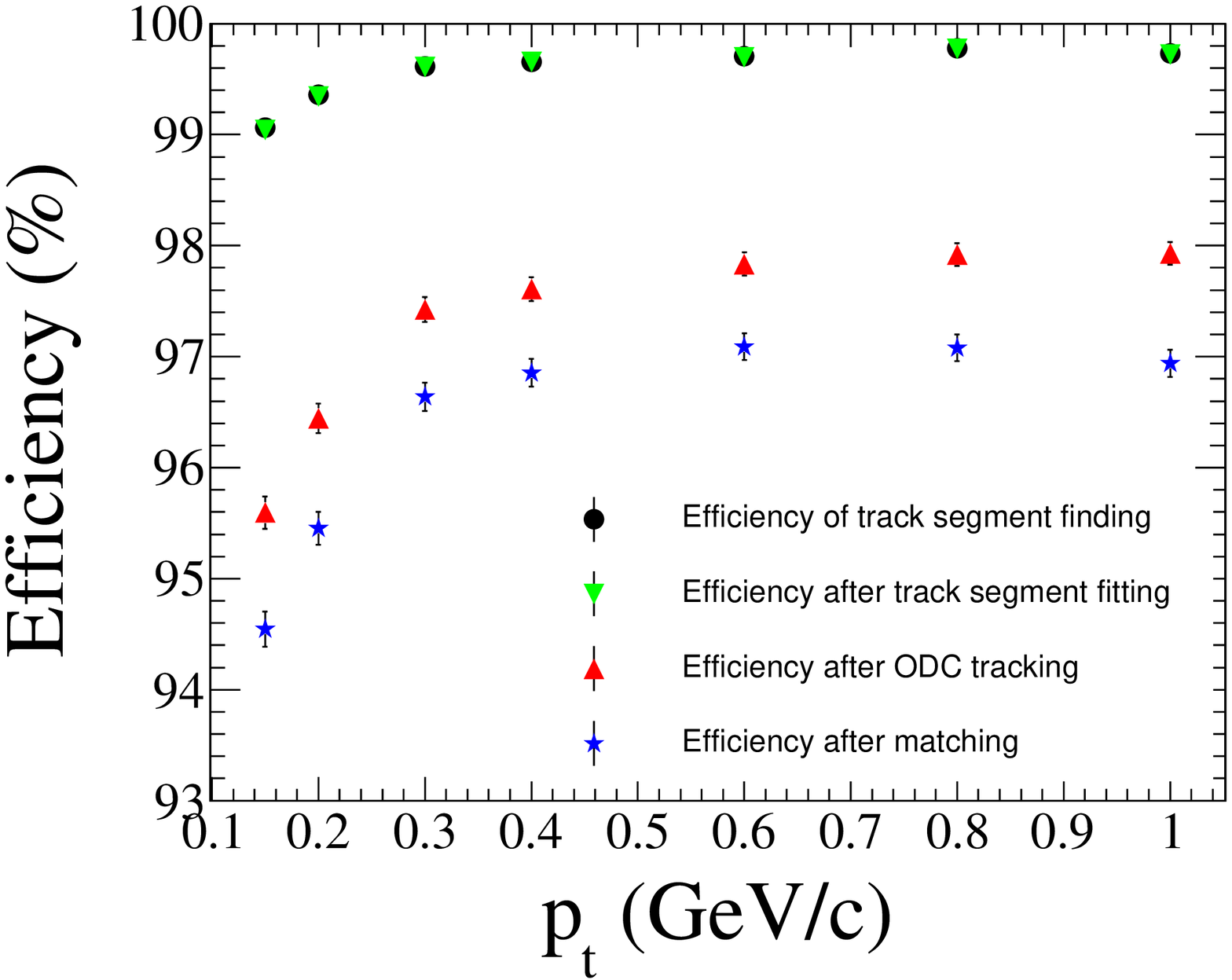}
\figcaption{\label{fig15} The absolute efficiency at different $p_t$ after the track segment finding in CGEM-IT, track segment fitting, ODC track finding and the track matching.}
\end{center}

In the study described here, the procedure of finding a complete track, in the tracking system consisting of the CGEM-IT
and the ODC, includes the track segment finding in CGEM-IT (black dots), track segment fitting (green inverted triangles), ODC track finding (red upright triangles) and the track matching (blue stars). These steps are proceeded in a sequence. The absolute efficiency after each step is shown in Fig.~\ref{fig15} at different $p_t$. The relative efficiency for the track matching is around $99\%$ and the total efficiency after the track matching ranges from $94.6\%$ to $97.0\%$ with some $p_t$ dependence.

\section{Summary and outlook}

As one candidate for the upgrade of the IDC at BESIII, the CGEM-IT was proposed and designed. The implementation and studies of the full simulation of the CGEM-IT and the track reconstruction are in progress. With a full simulated $\mu^\pm$ sample, two patterns are found on the relative difference in azimuth angle and in the position along the beam direction. These patterns are parameterized and used to select the cluster-combinations as track segments in CGEM-IT with an efficiency more than $99\%$. The quantitative description of the track segment is obtained by a helix fitting. The chi-square quantities evaluating the consistency between the track segments in CGEM-IT and the tracks in ODC are calculated and used in the track matching. Appropriate requirements on the chi-squares are chosen and the relative efficiency for the track matching is about $99\%$. A reasonable total efficiency, which is between  $94.6\%$ and $97\%$ depending on the transverse momentum of tracks,
 is obtained after the track segment finding with CGEM-IT, track segment fitting, track finding with ODC and the track matching.

To further improve the total track finding efficiency with the tracking system consisting of the CGEM-IT and the ODC,
one should optimize the track finding with ODC and develop a global track finding algorithm with both CGEM clusters
and ODC hits when it is hard to find good track segments in CGEM-IT or tracks in ODC.
Additional efforts should be devoted to the tracking of low momentum charged particles.
And all the charged track reconstruction algorithms should also be fine tuned for different particles ($e^{\pm}$, $\mu^{\pm}$, $\pi^{\pm}$, $K^{\pm}$, $p/\bar{p}$) with the existence of the backgrounds.


\begin{small}

\end{small}
\end{multicols}

\vspace{-1mm}
\centerline{\rule{80mm}{0.1pt}}
\vspace{2mm}

\begin{multicols}{2}

\end{multicols}

\clearpage
\end{document}